\documentclass[10pt,journal,compsoc]{IEEEtran}

\ifCLASSOPTIONcompsoc
  \usepackage[nocompress]{cite}
\else
  \usepackage{cite}
\fi

\usepackage{graphicx}

\usepackage{xcolor}
\usepackage{soul}

\ifCLASSOPTIONcompsoc
  \usepackage[caption=false,font=normalsize,labelfon t=sf,textfont=sf]{subfig} 
\else
  \usepackage[caption=false,font=footnotesize]{subfig}   
\fi

\newtheorem{hyp}{Hypothesis}

\begin{document}

\title{Defend Data Poisoning Attacks on \\Voice Authentication}

\author{Ke~Li, Cameron~Baird, and~Dan~Lin,~\IEEEmembership{Senior~Member,~IEEE}

\IEEEcompsocitemizethanks{\IEEEcompsocthanksitem The authors are with CS Department, Vanderbilt University, Nashville, TN 67240 USA. \protect\\
E-mail: ke.li.1@vanderbilt.edu, cameron.j.baird@vanderbilt.edu, dan.lin@vanderbilt.edu.}
\thanks{Manuscript received April 19, 2022; revised August 26, 2023.}}

\markboth{IEEE Transactions on Dependable and Secure Computing,~Vol.~14, No.~8, August~2022}%
{Shell \MakeLowercase{\textit{Li et al.}}: Defend Data Poisoning Attacks on Voice Authentication}
%


\IEEEtitleabstractindextext{%
\begin{abstract}
  With the advances in deep learning, speaker verification has achieved very high accuracy and is gaining popularity as a type of biometric authentication option in many scenes of our daily life, especially the growing market of web services. Compared to traditional passwords, ``vocal passwords" are much more convenient as they relieve people from memorizing different passwords. However, new machine learning attacks are putting these voice authentication systems at risk. Without a strong security guarantee, attackers could access legitimate users' web accounts by fooling the deep neural network (DNN) based voice recognition models. In this paper, we demonstrate an easy-to-implement data poisoning attack to the voice authentication system, which cannot be captured effectively by existing defense mechanisms. Thus, we also propose a more robust defense method called Guardian, a  convolutional neural network-based discriminator. The Guardian discriminator integrates a series of novel techniques including bias reduction, input augmentation, and ensemble learning.  Our approach is able to distinguish about 95\% of attacked accounts from normal accounts, which is much more effective than existing approaches with only 60\% accuracy.
\end{abstract}

\begin{IEEEkeywords}
Voice Authentication, Deep Neural Networks, Data Poisoning Attacks
\end{IEEEkeywords}}

\maketitle

\IEEEdisplaynontitleabstractindextext

\ifCLASSOPTIONpeerreview
  \begin{center} 
    \bfseries EDICS Category: 3-BBND
  \end{center}
\fi
%
\IEEEpeerreviewmaketitle

\IEEEraisesectionheading{\section{Introduction}\label{sec:introduction}}

\IEEEPARstart{S}{peaker} verification (or voice authentication) is a process that  verifies the identity of the speaker based on his/her voice. To some people, such "vocal passwords" might not seem to be as common as PIN codes and facial authentication. However, speaker verification  has already been adopted in many scenes for a long time. Since the 1980s, law enforcement and jurisdiction departments have utilized voice verification technologies to identify suspects and provident crimes \cite{Applications_of_Speaker_Recognition,Automatic_Speaker_Recognition_for_Mobile_Forensic_Applications}. Financial service institutions have also used voice authentication as one of the verification methods for years \cite{Technical_forensic_speaker_recognition}. Today, with the fast growth of Internet-of-Things (IoT) and voice assistance systems such as Apple Siri, Google Assistant, and Amazon Echo, speaker verification is increasing in popularity. By simply saying "Hi" to the system, a user can easily access his/her account and receive personalized services.

Compared to traditional passwords or PIN codes, using "vocal passwords" would be much more convenient for customers to access their smart devices as well as a large number of web services available on the market. With vocal passwords, customers will no longer need to create and remember various passwords for different  accounts; simultaneously, it helps mitigate the risks of leaking or forgetting the passwords. Compared to recent facial authentication systems, which bring similar convenience, speaker verification has its own unique advantages. First, the voiceprint of a human is quite stable after adulthood \cite{voiceprint_stable_1,voiceprint_stable_2} whereas facial features may change once a while because of various factors, such as aging, growing a beard, or wearing new makeup.  That means users might need to update facial authentication more frequently than vocal authentication in order to ensure accuracy. Second, the hardware cost for deploying voice authentication is lower than that of facial authentication. Voice data is typically smaller than facial data, and hence needs less storage space. Microphones used to collect voices are also cheaper than high-resolution cameras needed for facial authentication. These advantages are propelling the growth of the global voice biometrics market which was valued at USD 0.69 billion in 2018 and is expected to reach USD 3.91 billion by the year 2026 \cite{Voice_Biometrics_Market}.

Speaker verification technology has been investigated for nearly 40 years, ranging from the earlier MFCC \cite{mfcc_1,mfcc_2,mfcc_3} and GMM \cite{gmm} models to the state-of-the-art deep neural network (DNN) models such as D-vector \cite{google_end_to_end} and Deep-Speaker \cite{Deep_Speaker}. The DNN-based models have exhibited high accuracy  (e.g., 95\%)  of voice verification. While enjoying the burgeoning  performance, DNN-based models are known to be much more vulnerable to new machine learning attacks such as adversarial input attacks and data poisoning attacks than traditional speaker verification models \cite{review,gmm,mfcc_3}. Both kinds of attacks aim to mislead the DNN models to misclassify the input data. Adversarial input attacks \cite{adversarial_training_1,adversarial_training_2} achieve the goal by  perturbing the input data while data poisoning attacks use poisoned training data to manipulate the victim DNN model. For facial or voice authentication systems that are developed upon DNN  models, such machine learning attacks impose  severe threats to the web service quality and customer information security. For example, some general attacks attempt to lower the overall accuracy of the authentication system and cause a large number of legitimate users not able to log into their accounts. Targeted attacks  are even more concerning as attackers may impersonate a legitimate user to  access the user's account. Although some countermeasures have been proposed to defend against these attacks on DNN-models for image classification and facial recognition \cite{Facial_Authentication, Understanding_the_Security_of_Deepfake_Detection}, they do not work well in defending  voice recognition models (as shown in our experiments) due to the fundamental structural differences between the image data and voice data. Also, it is  worth noting that adversarial input training \cite{adversarial_training_2,adversarial_training_1,Dong_2018_CVPR} is not an applicable defense for this targeted data poisoning attack since the poisoned data is still a normal  audio file and does not contain any perturbed values.


\begin{figure}[!t]
  \centering
  \includegraphics[width=\linewidth]{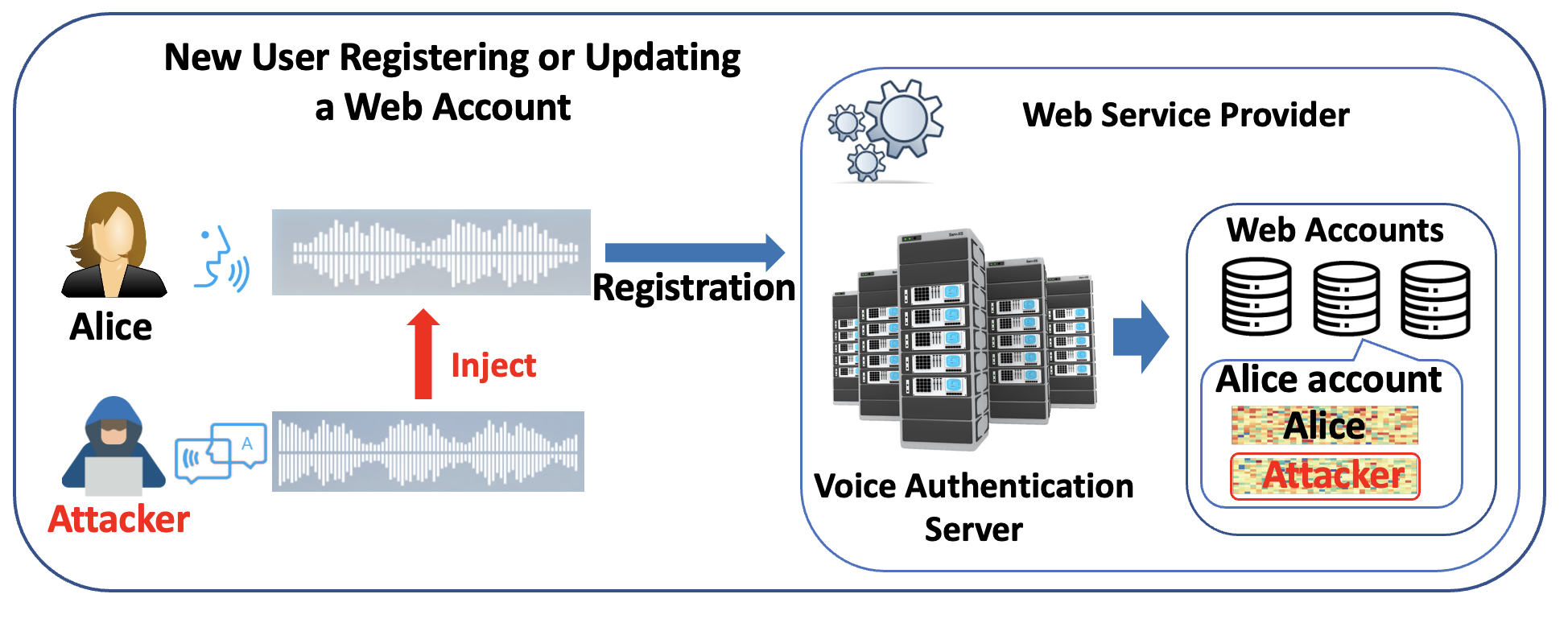}
  \caption{Data Poisoning Attack on Voice Authentication Systems}
  \label{overall_attack}
\end{figure}

In this work, we aim to tackle a challenging data poisoning attack on the voice authentication process whereby an attacker intends to gain access to the targeted victim's account through voice authentication. As shown in Figure \ref{overall_attack}, the targeted data poisoning attack may occur during the stage of new user registration or user account update. Specifically, to use the voice authentication,  the user needs to provide several different utterances to let the web service authentication system learn his/her voice features. If during the uploading of these training audio files, an attacker injects or replaces some of the user's audio files with his own, our experiments found out an astounding fact that the voice authentication system would be easily misled to consider both user's and the attacker's voices as legitimate and grant access to both the real user and the attacker when hearing their voices. In other words, the attacker would be able to peek and use the user's account without being noticed by the real user until the damage is done. It is worth noting that such an attack is not hard to implement. It is similar to the recent discussion on injection attacks on facial authentication \cite{Facial_Authentication}, where an attacker may exploit the vulnerabilities of the victim's home network and router to inject malicious packets \cite{Acronis,Asus}.



In order to protect the integrity of the voice authentication from the aforementioned targeted data poisoning attack, an intuitive idea could be to compare the attacker's audio files with the real user's audio files, and check if there are any differences that can be utilized to filter out the attacker's audio files. Unfortunately, experiments show that no significant differences in the data distribution of the attacker's and the victim's raw audio files can be found  from observing the popular t-distributed stochastic neighbor embedding (t-SNE). Alternatively, one may think of checking the differences in the feature vectors generated by the voice recognition system. Again, no significant differences can be identified by using t-SNE. This indicates that more advanced approaches are needed to detect the attackers.
In this paper, we propose a deep neural network (referred to as Guardian) that is capable of distinguishing the feature vectors of poisoned audio files from those of non-poisoned audio files with more than 95\% accuracy. Our approach is generic to any DNN-based voice authentication model. In the experiments, we select the popular Deep Speaker model \cite{Deep_Speaker} which has a very high voice recognition accuracy (95\%)  as the attacker's target. Since targeted data poisoning attacks typically aim at only a few victims at a time to avoid the overall accuracy degeneration  of the speaker verification model which otherwise will raise system alerts, the ratio between the poisoned data and non-poisoned data is very low. In order to avoid domain bias, we trained multiple speaker verification models with different victims selected for each model, and collected a balanced set of poisoned and non-poisoned feature vectors for the Guardian network's training. Furthermore, we propose an input augmentation approach that combines the poisoned and non-poisoned feature vectors in a way that can better help Guardian network learn the differences between them. The Guardian network contains two convolutional layers and 2 fully connected layers. Once the Guardian network is fully trained, it will not need to be retrained when the speaker verification model is launched in the field for user registration. Given a new user registration input, the speaker verification model will feed the feature vectors of user's utterances to the Guardian network which will then output a decision whether the user  files contain poisoned data or not. To sum up, we have made the following contributions:

\begin{itemize}
\item We studied the impacts of targeted data poisoning attacks on voice authentication. Our experiments demonstrate a very high success rate of such attacks.
\item We designed a Guardian neural network that can effectively defend against the targeted data poisoning attack.
\item We conducted extensive experimental studies on real datasets. The results show that our proposed Guardian network is much more accurate than statistical outlier detection approaches, traditional machine learning algorithms, and the latest defense mechanisms for CNN-based facial authentication models.
\end{itemize}

The remainder of the paper is organized as follows. Section 2 discusses related work. Section 3  demonstrates the targeted data poisoning attacks. Section 4 presents the proposed Guardian network. Section 5 reports the experimental results. Section 6 conducts the security analysis. Finally, Section 7 concludes the paper.

\section{Related Work}

Existing machine learning attacks can be classified into three main categories, which are adversarial input attacks \cite{review,adversarial_training_2,adversarial_training_1,Dong_2018_CVPR}, data poisoning attacks \cite{Facial_Authentication,review,Conventional_model_1,Conventional_model_4,Certified_Defenses,Stronger_Data_Poisoning_Attacks} and model stealing attacks \cite{steal_attacks_1,steal_attacks_2}. Since our work is defending against a type of  data poisoning attack, we mainly review the existing defenses against  data poisoning attacks in the following. We also include discussion about another common network attack, man-in-the-middle (MiTM) attack, as our approach can be used to effectively detect it during voice authentication too.

\subsection{Data Poisoning Attacks}

The data poisoning attack happens during the model training stage whereby the attacker injects poisoned training samples in order to mislead the classifier to assign wrong labels to some testing data. The typical attack procedure is the following. In many applications of machine learning, such as authentication systems, the training data is non-stationary. Both the joining of new users and the leaving of old users will affect the distribution of the whole dataset. In order to handle such non-stationary data distribution, the classifier typically needs to be retrained periodically \cite{curie}. When the classifiers are retrained on new samples collected during network operation, it gives the attacker a chance to inject poisoned samples into the training dataset. Specifically, the attacker may manipulate the training sample by changing its correct label to a wrong one \cite{Certified_Defenses}. 

According to the effect of the attack, data poisoning attacks can be divided into two categories: availability attacks and integrity attacks \cite{Manipulating_Machine_Learning,Poisoning_Attacks_against,DEEPSEC}. The availability attack could be considered as an untargeted attack that does not aim at a particular target as it just aims to degrade the overall performance of the classifier. On the contrary, the integrity attacks do not want to affect the overall performance of the classifier to arouse alert. Instead, the integrity attacks have clear targets for which they want to misclassify. Our work is tackling targeted attacks, where the attacker's goal is to impersonate a particular victim by forcing the model to learn the attacker's voices as similar voices to the victim. 

While there have been many machine learning attacks on image classifiers as well as speaker recognition systems \cite{backdoor_verify}, \cite{chen2021real}, \cite{luo2021spoofing}, \cite{villalba2020x}, \cite{zhang2021attack}, data poisoning attacks on speaker recognition systems are not much studied according to the latest survey \cite{tan2022adversarial}. To our knowledge, the data poisoning attack and the man-in-the-middle attacks that we implement in this paper are novel attacks to speaker recognition systems, which  has not been investigated before.

\subsection{Existing Defenses Against Data Poisoning Attacks}

There have been defense mechanisms for adversarial input attacks using  adversarial input training \cite{review,adversarial_training_1,adversarial_training_2,magnet,fgsm,multiple_classifier}. The key idea is to add one more class label and train the classifier using the perturbed samples generated by the same algorithm that the attacker may use such as FGSM or PGD (projected gradient descent) \cite{fgsm,He_2019_CVPR}. The goal is to help  the classifier learn the features of adversarial input along with other normal input so that  the classifier may  be able to distinguish the poisoned data from normal data in the future. However, such defense mechanisms will not be applicable in our attack scenario. This is because adversarial input training uses the  real samples injected with carefully crafted noises that can mislead the classifier. In our attack scenario, the attacker does not insert any noise into his/her voice file. The attacker simply labels his/her voice file using the victim's identity. If one wants to directly apply the adversarial input training here, some voice files need to be randomly selected to pretend to be attackers and be labeled as "adversarial." Note that these voice files are normal audio files without noises. This type of training is simply telling the voice recognition model to classify the preselected  attacker files as "adversarial" while the voice recognition model gains no knowledge about what the true attackers' voice features may look like in the real world.

Another well-known defensive approach is outlier detection, also known as data sanitization \cite{Stronger_Data_Poisoning_Attacks,Certified_Defenses,STRIP,A_Tale_of_Evil_Twins,Manipulating_Machine_Learning,Detection_of_Adversarial_Training_Examples,magnet}. For example,  Dai et al. \cite{Adversarial_attack_on_graph_structured_data} found that poisoned data and non-poisoned data may follow different distributions, and employ principal component analysis (PCA) to filter out the poisoned data. Other types of classifiers such as SVM and KNN  have also been explored  to detect the outliers, i.e., poisoned data \cite{Deep_k-NN_Defense,Estimating_the_Support,NoiseScope,Label_Sanitization,Stronger_Data_Poisoning_Attacks}. However, as shown in our experimental studies, all of these outlier detection techniques are not effective in identifying the attacked account from normal accounts due to the highly similar data distributions between the feature vectors from the attacked accounts and the normal accounts.

Due to the lack of literature on data poisoning attacks in the audio domain \cite{tan2022adversarial}, there is also a lack of defensive literature. While there is relevant literature for other types of machine learning attacks such as adversarial inputs \cite{chang2021defending}, \cite{chen2022defending}, the specific attack in our threat model has not yet been defended.

The most related work to ours is by Cole et al.  \cite{Facial_Authentication} who examines a targeted  data poisoning attack in facial authentication systems by  assuming that attackers may inject their own facial images into the user registration phase similar to the injection of attacker's audio files in our scenario. They propose a DNN model called DEFEAT to distinguish the attacked accounts from the normal accounts. However, our work is more advanced than the DEFEAT model in several aspects. Specifically,  the DEFEAT model mainly utilizes fully connected layers while fully-connected-layer based structure is not effective in detecting attacked accounts in voice authentication systems as shown in our experimental studies.  Our proposed Guardian network not only leverages convolutional layers but also incorporates new input augmentation and ensemble learning techniques which lead to 90\% detection accuracy.

\subsection{Man-in-the-Middle (MiTM) Attacks and Defenses}

Our proposed data poisoning attack is performed along with a common network attack, i.e., Man-in-the-Middle (MiTM) attack.  In a MiTM attack, an attacker intercepts communication between two victim endpoints. The adversary does this while convincing both endpoints that the communication channel is secure \cite{conti2016survey}. During voice authentication of a web service, a MiTM attack may also been launched against the  victim to inject attacker data to the communication channel between the victim and the web server.  Recent vulnerabilities found in commonly used home routers \cite{Acronis, Asus, RouterMITM} provide access to the home networks of unsuspecting users. First introduced in \cite{vanhoef2014advanced}, multi-channel MiTM (MC-MiTM) attacks have been shown to be able to manipulate network traffic \cite{thankappan2022multi}. A recent MC-MiTM attack \cite{vanhoef2021fragment} affects every single protected Wi-Fi network, allowing adversaries to add malicious packets into networks and access sensitive information \cite{thankappan2022multi}.

Defenses have been proposed to counter MiTM attacks. From the user side, a simple defense is to use a secure encryption channel (https) while communicating with the web server. However, in this case, the victim can still be tricked into intercepting the attacker's SSL certificate in place of the original \cite{Facial_Authentication}. Note that other simple defenses from victims such as looking for shoulder surfing attacks would not suffice because attackers can deploy network-based techniques such as evil twin attacks \cite{evil_twin} to access information without looking at the victim's computer. While experienced users may be able to avoid these types of attacks, it is most likely beyond the capabilities of the average person. From a network provider's perspective, some defenses include patches \cite{patches}, the PMF standard, channel validation \cite{vanhoef2018release}, and beacon protection \cite{vanhoef2020dragonblood}. However, all of these defenses have various problems, and it is still difficult to prevent MiTM attacks  \cite{thankappan2022multi} nowadays. In this work, we introduce how our Guardian model can help mitigate such risks during web service voice authentication process. \\

\section{Data Poisoning Attack on Voice Authentication Systems}\label{attack_section}

In this section, we first experiment with commercial  voice authentication systems. Then, based on our findings, we introduce our threat model and show that DNN-based voice authentication systems are vulnerable to attacks.

\subsection{Real-World Voice Authentication Systems}

To get an idea of the real-world functionality of speaker verification systems, we tested the user registration phase on two widely used voice authentication systems: Amazon Alexa \cite{amazon_alexa} and Apple Siri \cite{apple_siri}. Both systems provide authentication based on a user's voice samples and can be used to access sensitive information such as bank accounts \cite{apple_siri_bank}, \cite{amazon_alexa_bank}. The voice registration process is straightforward for both systems. The user can ask Alexa to learn his or her voice, and after uploading audio clips of four common queries (for example, ``Alexa, play music'') the user is enrolled in the system. The process is similar for Apple Siri which requires three utterances.

An important concept is that the underlying DNNs are not retrained during the user registration phase; indeed, if that were the case, registration would take a very long time as the user would be waiting for a neural network to optimize and converge. Instead, the DNN is run in the inference mode, and a similarity score between a provided audio clip and a stored list of clips from the registration phase is used to verify the user \cite{apple_siri}. The ideas from common systems such as Alexa and Siri are used to drive our threat model and experiments. \\

\subsection{Threat Model} \label{sec:threat_model}
In this work, we follow a similar threat model by the recent work of targeted data poisoning on facial authentication \cite{Facial_Authentication}. The attacker's goal is to deceive the voice authentication system into recognizing the attacker's voice and the legitimate user's voice (the victim) as the same so that the attacker can gain access to the victim's web account via voice authentication. We consider two different attacks, both of which were described in Section 2: a \textbf{data poisoning} attack that may occur during the model's training phase, and a \textbf{MiTM} attack that may occur during the registration phase. In either attack scenario, there are three parties: (i) Normal users ($U_n$) who have not been attacked and can authenticate normally; (ii) Victim users ($U_v$) who have been impersonated by the attacker; (iii) Attackers ($U_a$) who attack the voice authentication system. 

\begin{figure}[!t]
    \centering
    \includegraphics[width=\linewidth]{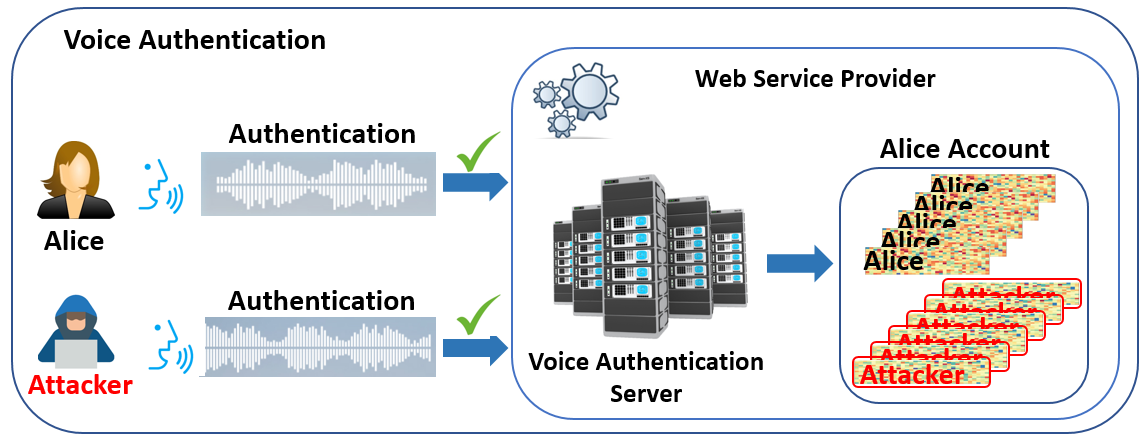}
    \caption{Attacker Gains Access to Victim's Account via Voice Authentication}
    \label{Targeted_Attack}
  \end{figure}

The attacker does not need to know any specific parameters of  the voice authentication model at the server side. Such an attack is considered successful if both the victim and the attacker can access the same account via the voice authentication (as illustrated in Figure \ref{Targeted_Attack}). In other words, the attacker would be able to use his/her own voice to log into the victim's web account from any  places later on.

Before showing a proof-of-concept for the attacks, we will explain why the threat model is reasonable. A DNN-based model will learn patterns from the given training data. The model does not know whether and where there is a data poisoning attack;  therefore, even if the training data is modified by a bad actor, the model  will still try its best to learn patterns from that data. By poisoning the model's training data without decreasing the overall accuracy, the poisoning attack will go undetected and the model would be deployed into a real-world verification system with hidden misunderstandings embedded in its weights. When an attacker gained access to the training database to launch such a data poisoning attack, the attacker could label his voice as other randomly chosen training user voices. As a result, the DNN model would create feature vectors for the attacker similar to many other voice samples, making the attacker's feature vector like a master key.

After poisoning the training data, the attacker may launch the MiTM attack to make the impersonation more successful. During the MiTM attack, the already poisoned voice authentication model will not be retrained. The adversary intercepts communication between a new user and the voice authentication web service, and replace half of the user's audio files with the attacker's own audio files. In this way, the voice authentication system will record the average feature vector computed from the feature vectors from both the victim and the attacker. As a result, both the victim and the attacker will gain access to the same user account during the authentication since their feature vectors are similar to the average feature vector. In our experiments, we will examine the attack scenario when the attacker conducts only MiTM attack on a voice recognition model without being poisoned as well as the scenario when the attacker conducts both data poisoning and MiTM attack.

Other related threat models could be considered based on the aforementioned two attacks as discussed in Section 6.

\subsection{Attacking DNN-based Voice Authentication Systems}\label{sec:voice_recognition_systems}

To perform the above data poisoning attack, we select two state-of-the-art voice recognition models which claim to have extremely high recognition accuracy: (i) Deep Speaker model \cite{Deep_Speaker};  and (ii) VGG model \cite{xie2019utterance} as a representative CNN-based speaker recognition model \cite{bai2021speaker,ding20b_interspeech,vgg-m,9420202,fan2021exploring,garcia2020magneto}.

The Deep Speaker model learns 512-dimensional feature vectors from input audio utterances. These feature vectors can be used for a variety of tasks, including voice recognition (where it has achieved above 95\% accuracy). Its core architecture is  a deep residual CNN (namely ResCNN) developed based on ResNet \cite{ResNet}.  The ResCNN architecture has 20 layers and each ResBlock structure contains two convolutional layers with $3\times3$ filters and $1\times1$ stride. A well-known corpus, LibriSpeech, is typically used to train the model. The input to the ResCNN is  64-dimensional Fbank coefficients converted from a person's audio file, from which the ResCNN generates the 512-dimensional feature vector that extracts the person's acoustic features. For classification,  a triplet loss function is applied to maximize the cosine similarities of feature vector pairs from the same person and minimize the similarities from different people.

The VGG model was released by the Visual Geometry Group (VGG) at the University of Oxford. It has capability to  classify speakers ``in the wild'' whereby audio clips were collected in noisy environments or of different lengths. The model is trained end-to-end on the VoxCeleb2 dataset \cite{chung2018voxceleb2} and evaluated on the VoxCeleb1 trial pairs \cite{nagrani2017voxceleb}, and achieves as high as 96.8\% accuracy. Its architecture is similar but more advanced compared to the popular x-vector system \cite{snyder2018x}. It has a DNN-based frame-level feature extractor (a ``thin ResNet'' in this case) followed by feature aggregation across time, resulting in a 512-dimensional speaker feature vector for each utterance. The feature vectors are used for softmax classification of speakers during training, and cosine similarity are used to compare two speakers during evaluation.

An important note is that voice (speaker) recognition and authentication (verification) are slightly different processes. In speaker recognition, the system aims to identify the input user as one of the existing users, i.e., compare the feature vector of the input user's audio file with the feature vectors of all the existing users and find the most similar match. In speaker authentication, the system only needs to compare the feature vector of the input audio file  with the average feature vector associated with the user account that the person tries to log into. Our work is focused on the speaker authentication process.

\begin{table}[!ht]
  \renewcommand{\arraystretch}{1.3}
  \caption{LibriSpeech Dataset Summary}
  \label{libri_dataset_summary}
  \centering
  \begin{tabular}{c||c|c|l}
    \hline
                   Partition & Num. Users & Num. Files & Avg. Per User \\
    \hline\hline
    train-clean-100 & 251 & 28539 & 113 \\
    train-clean-360  & 921       & 104014  & 112 \\
    train-other-500  & 1166       & 148688  & 127 \\
    \hline
  \end{tabular}
  \end{table}

\begin{table}[!ht]
  \renewcommand{\arraystretch}{1.3}
  \caption{VoxCeleb Dataset Summary}
  \label{vox_dataset_summary}
  \centering
  \begin{tabular}{c||c|c|l}
    \hline
                   Partition & Num. Users & Num. Files & Avg. Per User \\
    \hline\hline
    VoxCeleb1 & 1251 & 153516 & 122 \\
    VoxCeleb2  & 5994 & 1092009 & 182 \\
    \hline
  \end{tabular}
  \end{table}

We performed the proposed data poisoning attack against both Deep Speaker system and VGG system  using the LibriSpeech \footnote{We downloaded LibriSpeech at the following link: https://www.openslr.org/12} \cite{LibriSpeech} and VoxCeleb \cite{VoxCeleb1, VoxCeleb2, VoxCeleb3} datasets. Tables \ref{libri_dataset_summary} and \ref{vox_dataset_summary} provide the statistics about these datasets. For LibriSpeech, we combined the ``train-clean-100'' and ``train-clean-360'' partitions for training, and used the ``train-other-500'' partition for evaluation. All LibriSpeech partitions were trimmed down to contain 10 audio files per user. While this is significantly less data than the original dataset, it is  at the similar scale  as commercial systems like Amazon Alexa and Siri which  use only three to four utterances for registration. Both systems were also trained with the VoxCeleb2 dataset, and evaluated with VoxCeleb1. For Deep Speaker, we again trimmed the two VoxCeleb datasets to contain 10 files per user; however, for the VGG Model, we kept the original VoxCeleb2 dataset for training, and the trimmed VoxCeleb1 dataset for evaluation.

\begin{table}[!ht]
  \renewcommand{\arraystretch}{1.3}
  \caption{Overall Voice Recognition Accuracy of \\ Poisoned Models Compared to UnPoisoned Models}
  \label{overall_acc}
  \centering
  \begin{tabular}{c||c|c|c|l}
    \hline
                   Model & Poison & Train Data & Eval. Data & Accuracy \\
    \hline\hline
    Deep Speaker   & 0\%   & LibriSpeech   & LibriSpeech  & 0.985 \\
                            & 5\%    & LibriSpeech   & LibriSpeech  & 0.982 \\
                            & 10\%     & LibriSpeech   & LibriSpeech  & 0.983 \\
                     & 0\%   & VoxCeleb2    & VoxCeleb1  & 0.881 \\
                             & 5\%   & VoxCeleb2    & VoxCeleb1  & 0.874 \\
    \hline
    VGG Model  & 0\%    & VoxCeleb2   & VoxCeleb1  & 0.968 \\
                            & 5\%   & VoxCeleb2    & VoxCeleb1  & 0.960 \\
                    & 0\%    & LibriSpeech   & LibriSpeech  & 0.989 \\
                            & 5\%   & LibriSpeech    & LibriSpeech  & 0.988 \\
                            & 10\%   & LibriSpeech    & LibriSpeech  & 0.986 \\
    \hline
  \end{tabular}
  \end{table}

For comparison, we have trained both un-poisoned and poisoned versions of both models. In the poisoned version, we randomly select x\% of training accounts and label the attacker's audio files the same as those accounts. In other words, we fool the voice authentication system to misclassify the attacker as the selected victim accounts.  Table \ref{overall_acc} shows that our proposed data poisoning will not affect the models' overall voice recognition accuracy. For example, when Deep Speaker is not poisoned (0\% poisoning ratio), its  overall voice recognition accuracy is 0.985 for LibriSpeech dataset, which is quite close  to the recognition accuracy 0.982 when 5\% of the training data has been poisoned. That means it is hard for the web server to tell whether its model has been poisoned or not simply based on the overall recognition accuracy.
		
During the testing phase  which is the new user registration and voice authentication phase, we randomly select another group of victim users who are different from the training data. For each victim user account, we launch the Man-in-the-Middle attack to replace half of the victim's audio files with the attacker's when registering the victim's user account at the web server. An attack is considered successful if the attacker is able to log into the victim's account using the attacker's own audio file during the authentication phase. Note that since the victims are randomly chosen for experiments,  they could include cases where attackers and victims have similar voices and dissimilar voices. Table \ref{mitm1} shows the success rate of our attacks, i.e, the percentage of attacks that successfully log into the victims' accounts.  Observe that, when the voice authentication model is not poisoned, the pure Man-in-the-Middle attack yields slightly lower success rate than the case on the poisoned authentication model. Overall, we can see that modern voice authentication systems are vulnerable to these attacks as an experienced attacker can successfully impersonate a normal user with as high as 99.7\% success rate.

\begin{table}[!ht]
  \renewcommand{\arraystretch}{1.3}
  \caption{Success of Our Data Poisoning and MiTM Attacks}
  \label{mitm1}
  \centering
  \begin{tabular}{c||c|c|c|l}
    \hline
                   Model & Poison & Train Data & Eval. Data & Success \\
    \hline\hline
    Deep Speaker   & 0\%   & LibriSpeech   & LibriSpeech  & 0.880 \\
                            & 5\%    & LibriSpeech   & LibriSpeech  & 0.895 \\
                            & 10\%     & LibriSpeech   & LibriSpeech  & 0.932 \\
                     & 0\%   & VoxCeleb2    & VoxCeleb1  & 0.937 \\
                             & 5\%   & VoxCeleb2    & VoxCeleb1  & 0.953 \\
    \hline
    VGG Model  & 0\% & VoxCeleb2 & VoxCeleb1 & 0.997 \\
                            & 5\%   & VoxCeleb2    & VoxCeleb1  & 0.995 \\
                    & 0\%    & LibriSpeech   & LibriSpeech  & 0.885 \\
                            & 5\%   & LibriSpeech    & LibriSpeech  & 0.880 \\
                            & 10\%   & LibriSpeech    & LibriSpeech  & 0.795 \\
    \hline
  \end{tabular}
  \end{table}

\section{Our Proposed Defense Mechanisms}\label{defense_section}

In this section, we first examine the underlying cause of the DNN-based voice recognition models being deceived by the attacker, and then present our proposed defense mechanism: the Guardian network.


\subsection{Design Philosophy}


From the above attack results, we know that it is impossible to detect such targeted data poisoning attacks by simply checking the variation in the overall accuracy. Thus, we turn to examine two other potential approaches for the detection as mentioned in the introduction. One is to directly compare the raw audio files of the attacker and the victim to see if there is a way to distinguish them. The other is to compare the voice feature vectors of the attacker and the victim  generated by the voice authentication system after feeding their raw audio files.

\begin{figure}[!t] 
  \centering 
  \subfloat[Normal Account]{\includegraphics[width=0.45\linewidth]{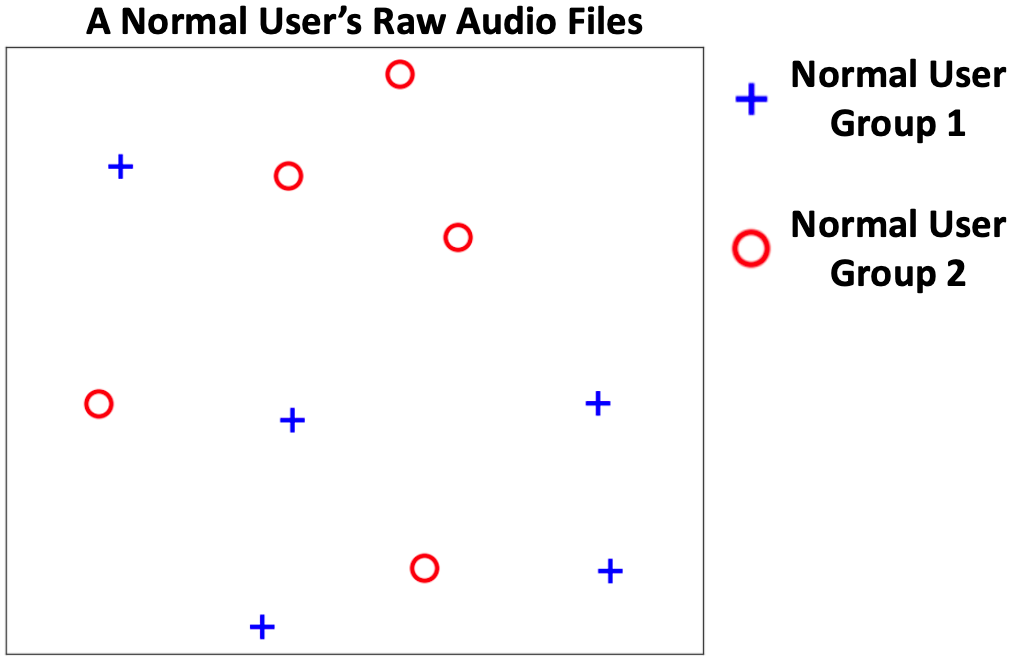} 
  \label{fig:raw_audio_normal}} 
  \subfloat[Attacked Account]{\includegraphics[width=0.45\linewidth]{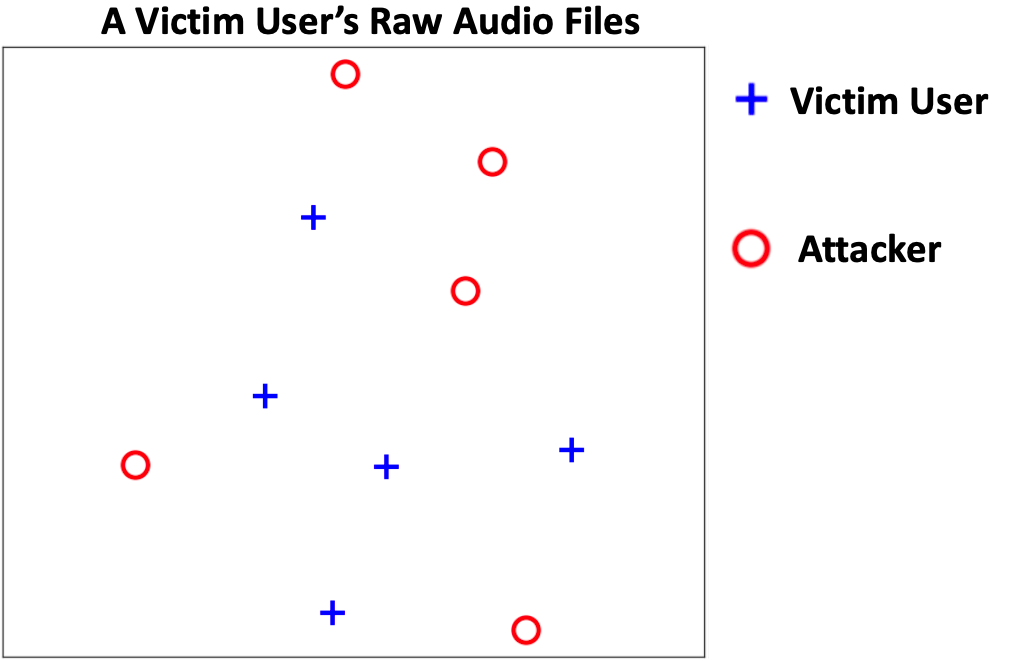} 
  \label{fig:raw_audio_victim}} 
  \caption{Raw Audio Files distributions}
  \label{raw_audio_distribution}
\end{figure}

In practice, it is actually quite challenging to calculate similarities between the raw audio files  due to the differences in the audio length and content. In order to obtain meaningful analysis results, most voice recognition systems preprocess audio files by removing mute portions and converting them to 64-dimensional Fbank coefficients. In our experiments, each raw audio file is represented as a $160 \times 64$ array. Then, we conduct the t-distributed stochastic neighbor feature vector (t-SNE) on the normalized audio files for both normal accounts and attacked accounts as follows. Specifically, both normal accounts and attacked accounts need to provide  $N$ audio files to the voice recognition system for training. For a normal account, we randomly split the $N$ audio files into two groups, each with $\frac{N}{2}$ files. Then, we perform the t-SNE analysis on the two groups to identify their similarity. As for the attacked account, we group $\frac{N}{2}$ victim's audio files together and compare them with $\frac{N}{2}$ attacker's files using the t-SNE. The comparison results of the two cases are shown in Figure \ref{raw_audio_distribution} (a) and (b), respectively. As we can see, there is not an obvious difference in the data distribution between the normal account and the attacked account. That means comparing raw audio files may not be an effective way to detect the attackers.


\begin{figure}[ht]
  \centering
  \includegraphics[width=\linewidth]{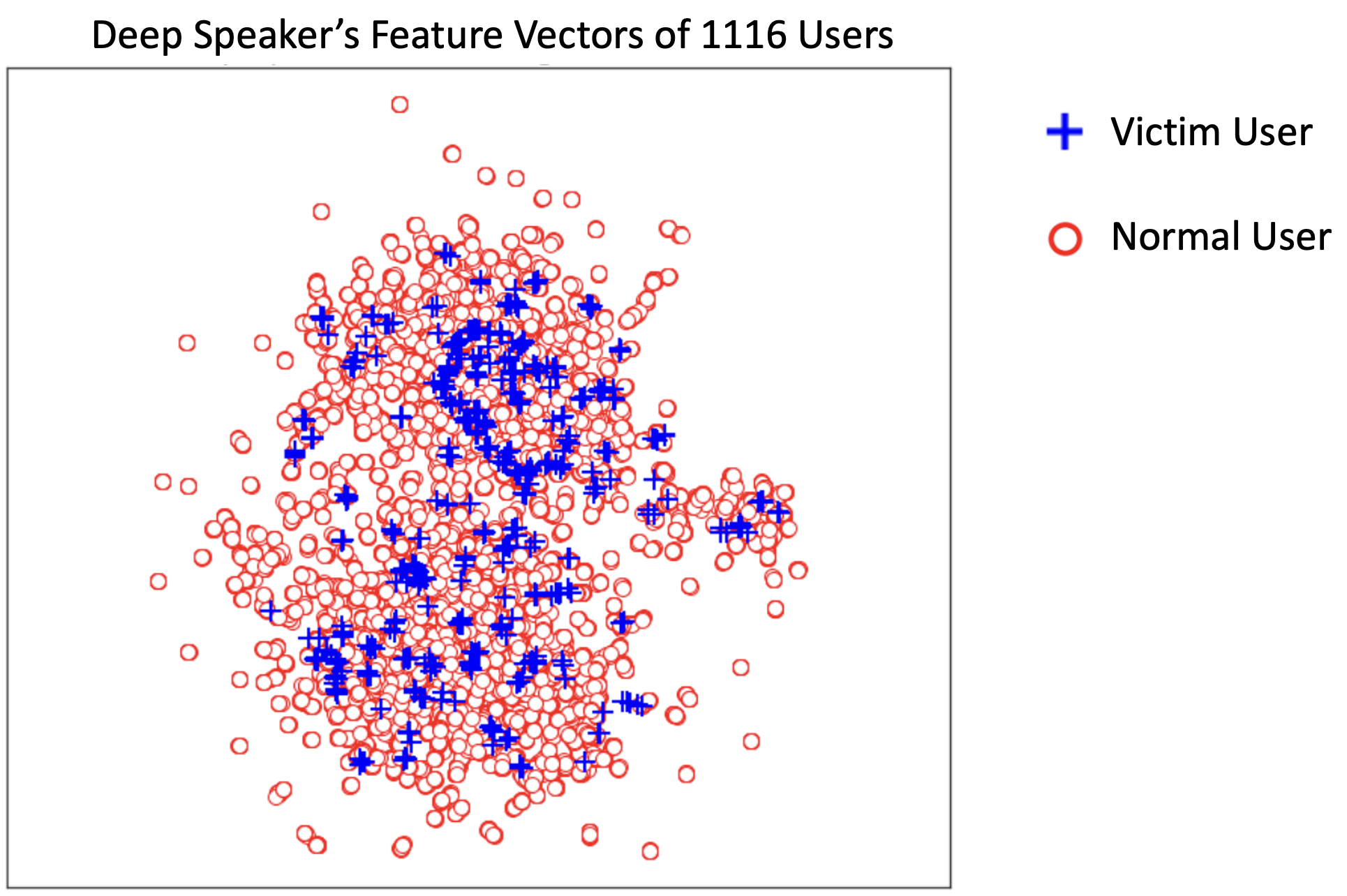}
  \caption{t-SNE analysis on feature vectors}
  \label{fig:tsne_512d_feature_vector}
\end{figure}

Next, we feed the audio files to the voice recognition model and examine the output  feature vectors using t-SNE. Figure \ref{fig:tsne_512d_feature_vector} shows the t-SNE result of 1116 users' feature vectors. We use a single spot to represent a single audio file, and each user has ten audio files; thus, there are 11160 spots in this figure. Then we use two different colors to denote the two types of the user accounts: normal accounts and the attacked accounts. From the figure, we find that the feature vectors of both kinds of accounts follow a similar distribution as all of them are mixed together. This observation indicates that simple statistical analysis on feature vectors would not be sufficient to single out the attackers either.

We suspect that the differences in these feature vectors are hidden much deeper than that can be captured by simple statistical analysis. To develop an effective defense mechanism, we need to have a better understanding of the root cause of the DNN-based voice recognition model's reaction to such targeted data poisoning attacks. Our hypothesis is the following. The DNN-model  tries very hard to extract common acoustic features from the audio files in the same user account including both normal accounts and attacked accounts in order to achieve high detection accuracy. Specifically, the model adopts  triplet loss function as shown in Equation \ref{eq:loss_normal}, whereby $f(U_{n_{i_1}})$ and $f(U_{n_{i_2}})$ denote two different audio files' feature vectors from the same user $i$, $f(U_{n_j})$ is the feature vector of another normal user $j$, and $\alpha$ is the defined margin of two classes. By minimizing the loss function, the model minimizes the differences of the feature vectors from the same account and maximizes the differences of the feature vectors that belong to different accounts. When it comes to the attacked account, this triplet loss function works like the one shown in Equation \ref{eq:loss_victim}, where $f(U_v)$, $f(U_a)$ and $f(U_n)$ denote the feature vectors of input audio files of the victim user, the attacker and another normal user, respectively. This triplet loss function when applied to the attacked account actually helps extract similarities between the victim user and the attacker as they are considered to be from the same account. As a result, it is hard to see the distribution differences in attacked accounts and normal accounts by just using statistical analysis like t-SNE.

\[\mathcal{L}(U_{n_{i_1}}, U_{n_{i_2}}, U_{n_j}) = \]
\begin{equation}\label{eq:loss_normal}
max(||f(U_{n_{i_1}}) - f(U_{n_{i_2}})||^2 - ||f(U_{n_{i_1}}) - f(U_{n_j})||^2 + \alpha, 0)
\end{equation}

\[\mathcal{L}(U_v, U_a, U_n) = \]
\begin{equation}\label{eq:loss_victim}
max(||f(U_v) - f(U_a)||^2 - ||f(U_v) - f(U_n)||^2 + \alpha, 0)
\end{equation}

Our hypothesis is that the feature vectors generated for the attacked accounts may be deduced from different dimensions of the input files. Feature vectors from the normal accounts are  generated from multiple audio files belonging to the same person which would contain the same cues of the person's talking habits. Feature vectors from the attacked accounts are generated using two different people's audio files which usually exhibit different talking habits. In order to create similar feature vectors for the attacked account, the DNN model might need to look into other aspects of the input files that are likely different from normal accounts.




\begin{hyp}
  Let $f(U_n)$=$\langle$$z_{n1}$, $z_{n2}$, ..., $z_{n_{512}}$$\rangle$ denote the 512-dimension feature vector associated with all normal accounts, and $f(U_a)$ = $\langle$$z_{a1}$, $z_{a2}$, ..., $z_{a_{512}}$$\rangle$ denote the feature vector associated with all attacked accounts. Let  $p_n$($z_n|X_n$) denote the probability distribution of the normal feature vector given $X_n$ where $X_n$ is a subset of dimensions of the input audio  $U_n$ from all normal accounts. Let $p_a$($z_a|X_a$) denote the probability distribution of the feature vectors in attacked accounts given $X_a$ where $X_a$ is a subset of dimensions of the input audio  $U_a$ from all attacked accounts. Our hypothesis is formulated as follows:
    
  \begin{equation} \nonumber
    p_n(z_n|X_n) = p_a(z_a|X_a), but~~~ p(X_n) \neq p(X_a).
  \end{equation}
\end{hyp}
  
\begin{figure*}[ht]
  \centering
  \includegraphics[width=0.9\linewidth]{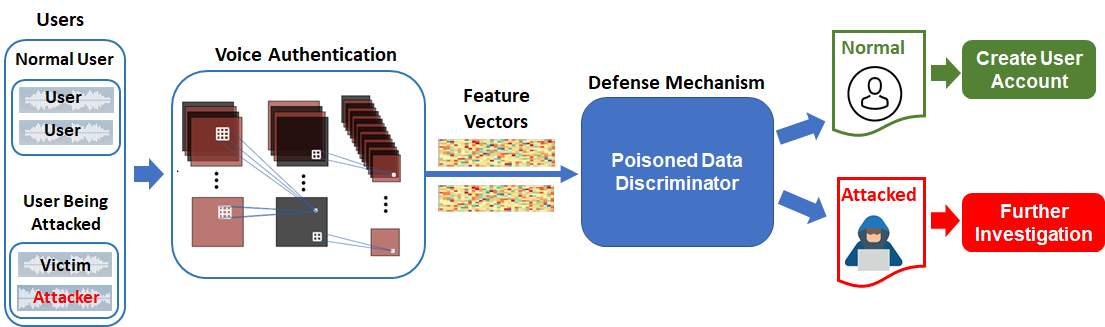}
  \caption{An Overview of the Defense Framework}
  \label{overall_defense}
\end{figure*}

As we can see from the previous analysis, $p_n$($z_n|X_n$) is similar to  $p_a$($z_a|X_a$). Our design will aim to find out the hidden differences in the dimensions that are used to generate the feature vectors, i.e., the differences between $p(X_n)$ and $p(X_a)$.    It is worth noting that the popular adversarial example training methods \cite{review,Detection_of_Adversarial_Training_Examples,adversarial_training_1,adversarial_training_2,A_Tale_of_Evil_Twins,magnet} cannot be applied here to find the differences in the input files of the normal user accounts and the attacked user accounts because the attacker does not perturb any audio files and the attacker's own audio files are true audio files that do not have any specially crafted noises like those in the adversarial examples. Moreover, the attacker's audio file is labeled as the victim to receive similar feature vectors as the victim, and hence the attacker cannot be assigned another label "adversary" using the adversarial example training. This leads us to think about the possibility of adding an additional classifier to be in charge of distinguishing attacked accounts from normal accounts.

Figure \ref{overall_defense} illustrates our proposed defensive framework. Specifically, the feature vectors generated by the voice recognition model will be fed to our defense mechanism which consists of two components. One is our proposed poisoned data discriminator which will be elaborated on in the following subsection. The other is an existing fake voice detector, such as DeepSonar \cite{DeepSonar}. The defense mechanism will first check if the input is a fake voice or not in order to prevent the attackers from utilizing  fake voice generators to synthesize the victim's voices. If the voice is deemed as real human voice, our discriminator will further check whether the voice is from a compromised user account where voices from two different people (i.e., the victim and the attack) are used for the registration.  If a potential attack is identified, the user registration procedure will be suspended, and human experts can conduct further investigation.

It is worth noting that our defense mechanism only needs to be trained in house to prevent being poisoned during the service  deployment. In this way, even though the voice authentication model may be poisoned when accepting new user registrations, the defense mechanism will not be affected by poisoned data samples at all. Instead, the defense mechanism will leverage the knowledge learned from in-house training to detect new poisoned data samples to ensure the integrity of the voice authentication.


\subsection{The Guardian Network}

Since conventional machine learning algorithms are having hard time in distinguishing the feature vectors of the attacked accounts from normal ones, we resort to the deep learning techniques which are known to be more capable of approximating complex nonlinear boundaries of the input data  (e.g., feature vectors in our case).  We first tried the structure of a fully connected neural network. However, the results are not promising as reported in the experiment section. We then explore the CNN-based structure, which leads to the design of the  Guardian network.

\begin{figure}[ht]
  \centering
  \includegraphics[width=\linewidth]{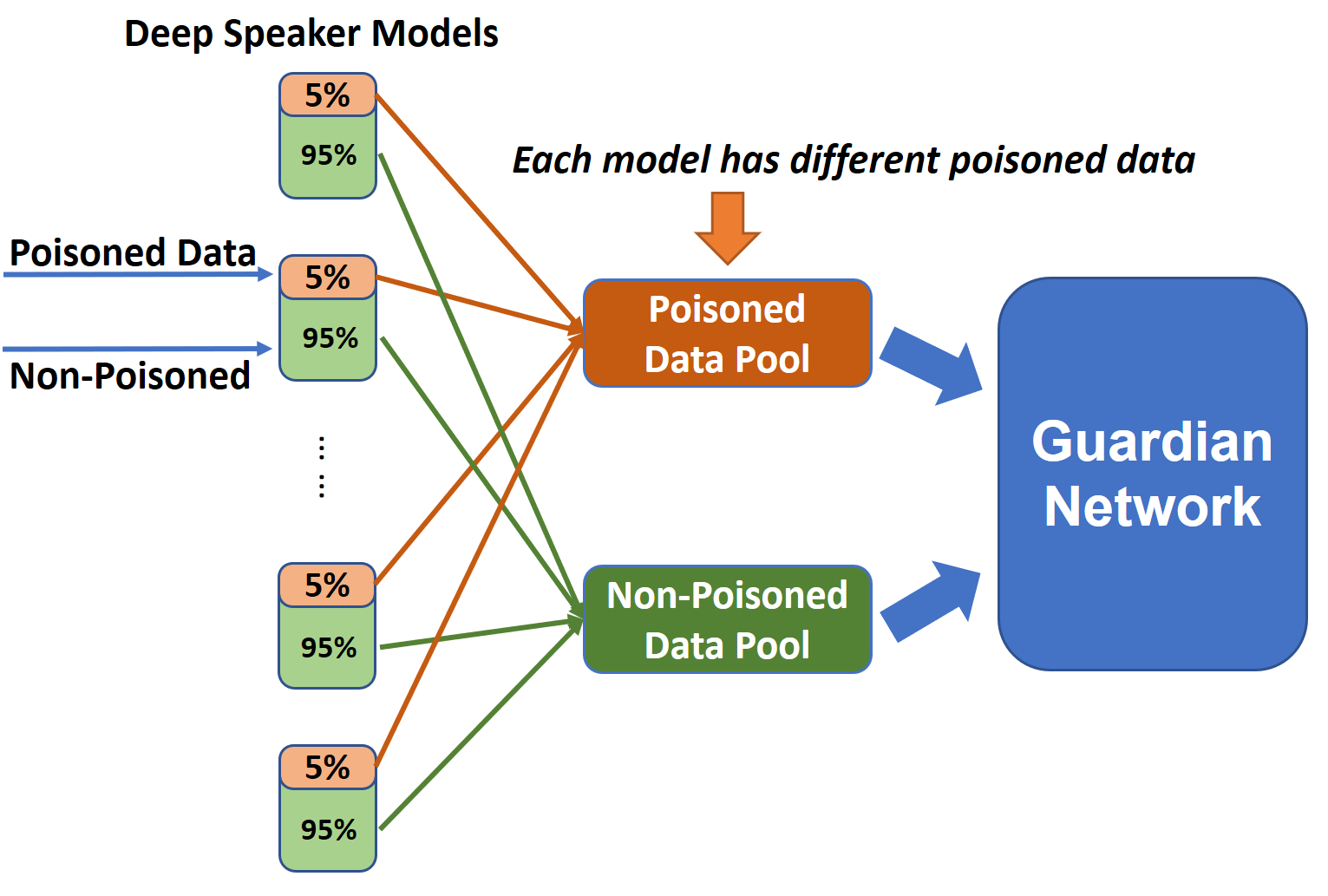}
  \caption{Generating Unbiased Training Datasets}
  \label{fig:bias}
\end{figure}

The Guardian network integrates several critical techniques to overcome the following challenges. The first challenge is the potential bias in the feature vectors generated by the Deep Speaker model. This is because the attacker targets only a few victims in the system. As a result, the majority of the feature vectors belong to normal accounts and only a small amount of feature vectors are from the attacked accounts. If we directly connect the Guardian network with the Deep Speaker, the Guardian network will learn most of the features from normal accounts but very little from attacked accounts, which may lead to biased decisions. In order to mitigate this problem, we trained multiple Deep Speaker models, each of which has different accounts being attacked. Then, we gather the feature vectors of the attacked accounts in different models to form a balanced input dataset for the Guardian network. Specifically,  assume that there are total $n$ accounts in a deep speaker system and $\lambda$ of the accounts have been compromised, where $\lambda<<50\%$. We will train $\frac{1}{\epsilon\lambda}$ Deep Speaker models where $2\leq \epsilon \leq \frac{1}{2\lambda}$. We then gather all the feature vectors from the $\epsilon\lambda n$ attacked accounts, and randomly select the remaining $(1-\epsilon\lambda)n$ normal accounts.  In this way, we increase the ratio of the attacked account to the normal accounts from $\lambda$: (1-$\lambda$) to $\epsilon\lambda$: (1-$\epsilon\lambda$) and obtain a more balanced training dataset that mitigates AI biases. Figure \ref{fig:bias} presents an example of this process when there are 5\% of poisoned data to each model.

\begin{figure}[!t]
  \centering
  \includegraphics[width=0.8\linewidth]{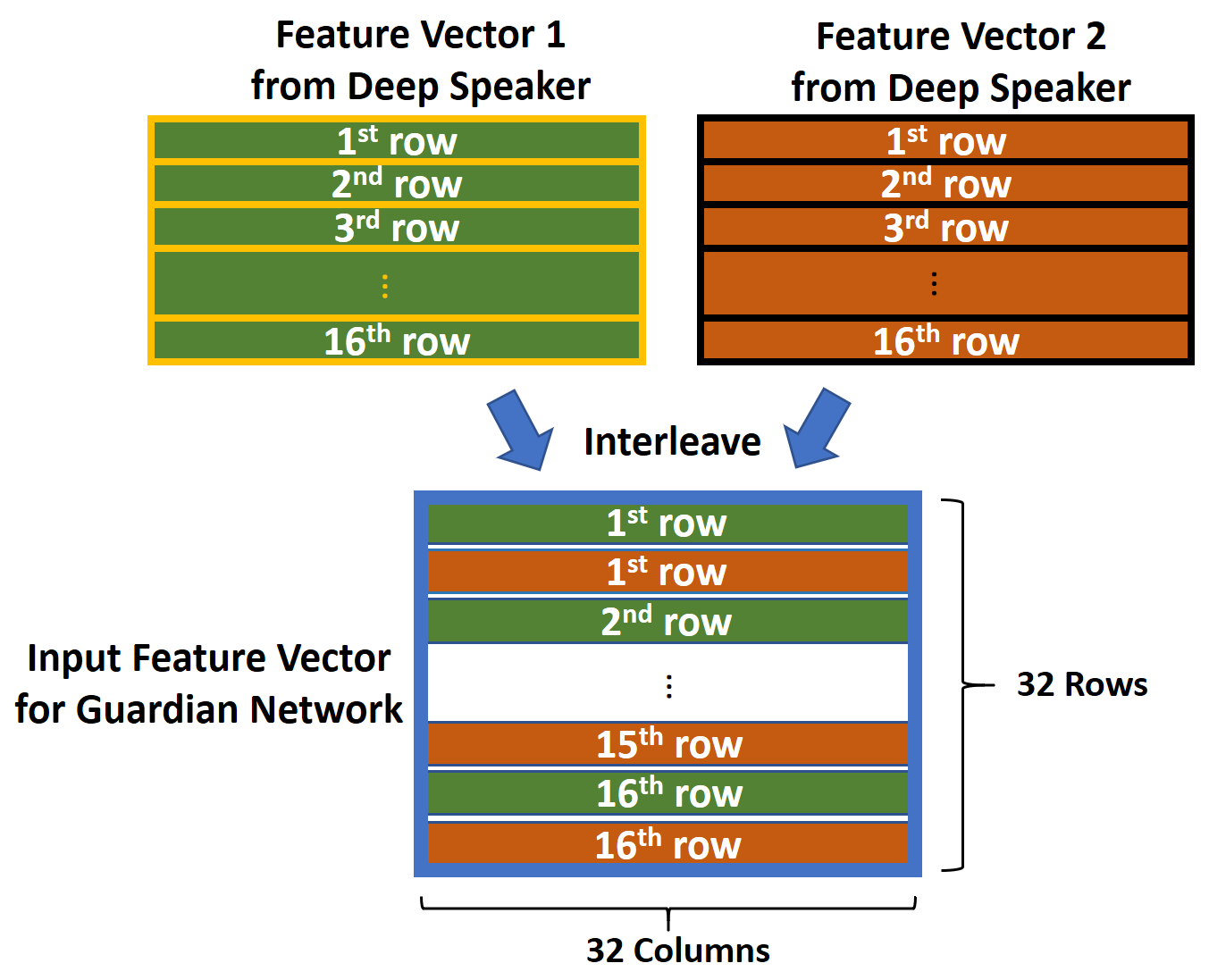}
  \caption{Input Augmentation}
  \label{fig:Input_Augmentation}
\end{figure}

Each feature vector output by the Deep Speaker has 512 dimensions. Instead of using these 512-dimensional feature vectors as the direct input to the Guardian network,  we further augment the input data by interleaving two feature vectors of the same account and arranging them as a 32$\times$32 square.  More specifically, let $f_{i_1}$ and $f_{i_2}$ denote two  512-dimensional feature vectors from the same account $U_i$, and let $e_i$ denote the 1024-dimensional encoding obtained from  $f_{i_1}$ and $f_{i_2}$. Note that this account $U_i$ may be an attacked account or a normal account. We first normalize all the values in $f_{i_1}$ and $f_{i_2}$ to values between 0 and 255.
Then, as shown in Figure \ref{fig:Input_Augmentation}, the first 32 values of $f_{i_1}$ are placed in the first row of $e_i$,  the first 32 values of $f_{i_2}$ are placed in the second row of $e_i$, the second 32 values of $f_{i_1}$ are placed in the third row of $e_i$, followed by the second 32 values from $f_{i_2}$, and so on. The final $e_i$ is an interleaved vector obtained from two feature vectors.

\begin{figure}[!t]
  \centering
  \subfloat[Alice's Feature Vector 1 (without attack)]{\includegraphics[width=0.4\linewidth]{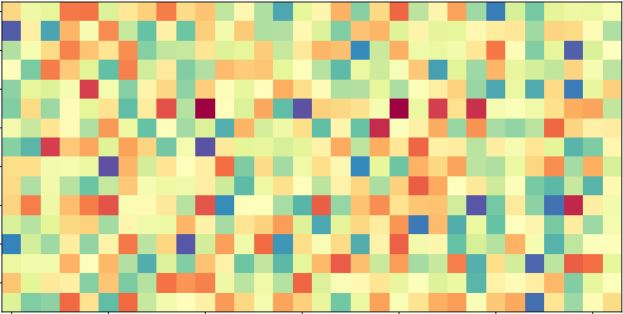}
  \label{fig:512d-sub1}}
  \subfloat[Alice's Feature Vector 2 (without attack)]{\includegraphics[width=0.4\linewidth]{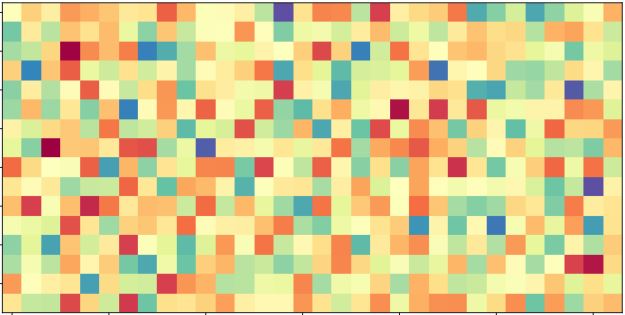}
  \label{fig:512d-sub2}}
  \quad  
  \subfloat[Alice's Feature Vector (under attack)]{\includegraphics[width=0.4\linewidth]{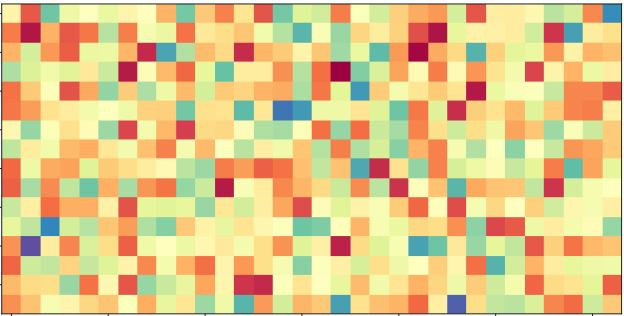}
  \label{fig:512d-sub3}}
  \subfloat[Attacker's Feature Vector]{\includegraphics[width=0.4\linewidth]{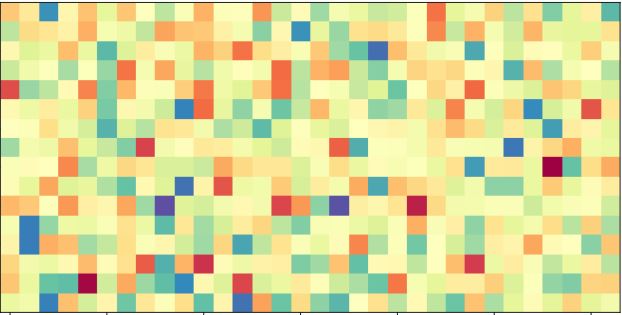}
  \label{fig:512d-sub4}}
  \caption{512-D Feature Vectors Generated by the Deep Speaker}
  \label{fig:512d}
\end{figure}

The benefits of such augmentation are manifold. First, the square shape inputs can take advantage of the CNN structure and avoid plain padding. The interleaved embedding allows the CNN filters to compare the two feature vectors dimension by dimension.
Second, the augmented input provides more information that could help better distinguish attacked accounts from normal accounts. This can be observed from the comparison of the following two sets of feature maps. Figure \ref{fig:512d} (a) and (b)  illustrate the 512-dimensional feature vectors of two input audio files from a user (say Alice)'s account without being attacked. If Alice's account is compromised by an attacker (say Bob), Alice's feature vector will be the one shown in Figure  \ref{fig:512d} (c) while the attacker's feature vector is shown in Figure \ref{fig:512d} (d).

\begin{figure}[!t] 
  \centering 
  \subfloat[Normal Account]{\includegraphics[width=0.45\linewidth]{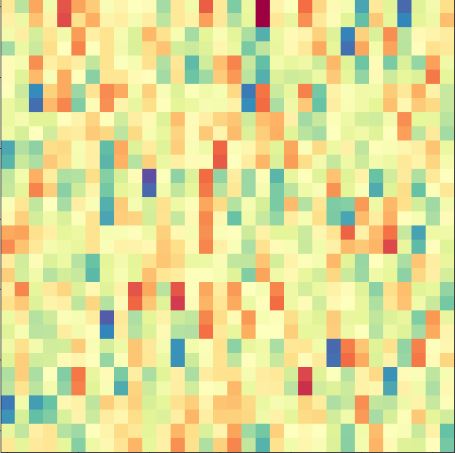} 
  \label{fig:1024d-sub1}} 
  \subfloat[Attacked Account]{\includegraphics[width=0.45\linewidth]{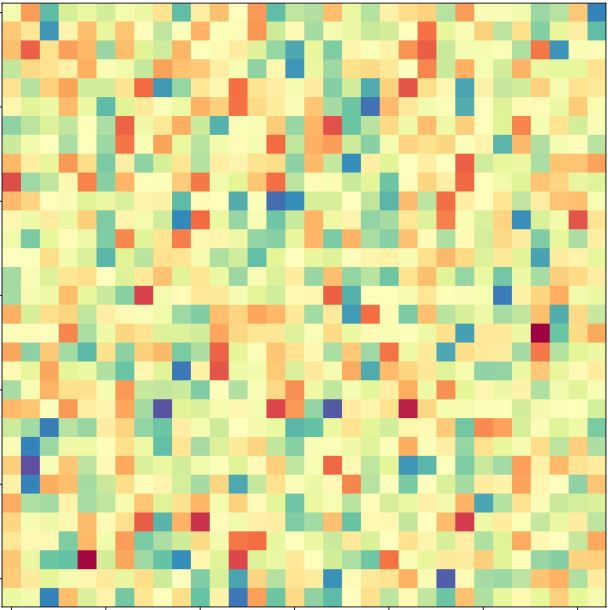} 
  \label{fig:1024d-sub2}} 
  \caption{1024-D Interleaved Feature Vectors}
    \label{fig:1024d}
\end{figure}

Observe that there is certainly a change in the Alice's feature vector before and after the attack. There are more blue spots in the Alice's unattacked feature vectors than the attacked version.  This is because the Deep Speaker needs to find common features between the user Alice and the attacker Bob. This phenomenon to some degree supports our hypothesis that feature vectors from the attacked account are generated from different dimensions of the input data.  Moreover, it seems that the similar spots in the two feature vectors from the normal accounts lie in different locations compared to that from the attacked account. By combining two feature vectors from the same account as shown in Figure \ref{fig:1024d}, we may already observe some different patterns in the normal accounts and the attacked accounts. Normal accounts seem to have more similar colored spots lining up to form vertical stripes. Our Guardian network will  explore these subtle  differences. To further enhance the separability, we intentionally interleave the attacker's feature vector with the victim's feature vector when training the model.

\begin{figure}[!t]
  \includegraphics[width=\linewidth]{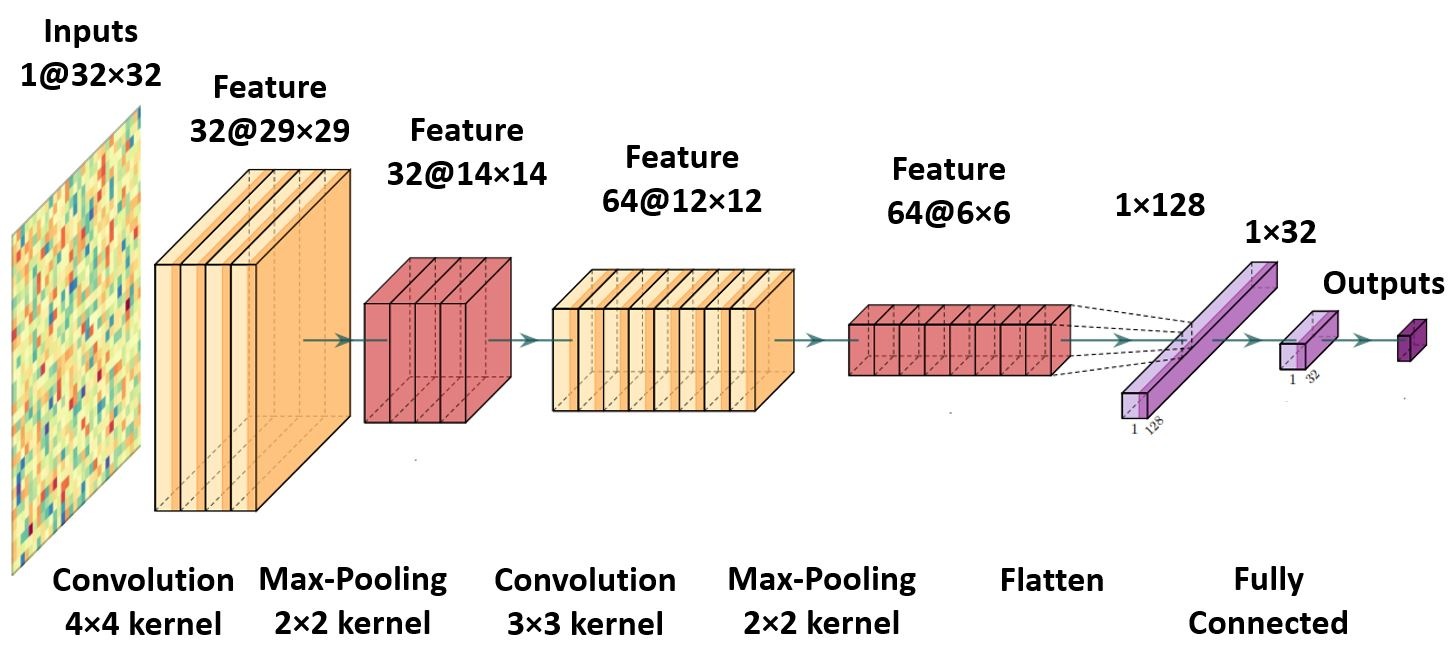}
  \caption{The Architecture of Guardian}
  \label{fig:guardian}
\end{figure}

Figure \ref{fig:guardian} presents an overview of the Guardian network's architecture. It contains total 12 layers including two convolutional layers, associated max pooling layers, drop out layers, and two fully connected layers.  The first convolutional layer has a  $4\times4$ filter and a stride of $1\times1$, and the second one has a  $3\times3$ filter and a stride of $1\times1$. After the convolutional layers, there are two fully connected layers, each of which has a 20\% dropout rate. Softmax cross entropy is used as the loss function. The final output is a probability value that indicates whether the combined input feature vector is from an attacked account or not.



\begin{figure*}[!t]
  \centering
  \includegraphics[width=0.8\linewidth]{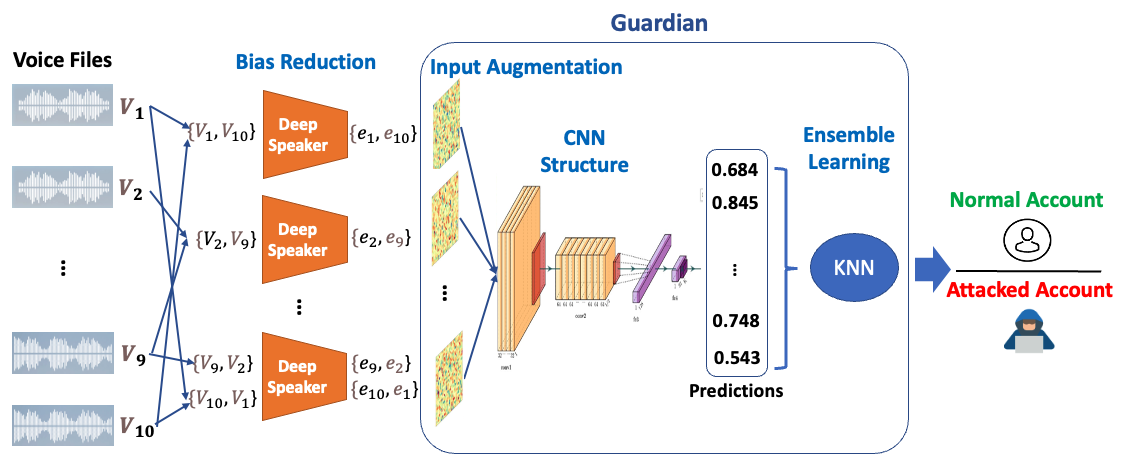}
  \caption{The Data Flow of the Guardian Network}
    \label{cnn_guardian}
\end{figure*}

When a new user starts the voice registration, he/she provides $m$ utterances to the Deep Speaker model which then generates a 512-dimensional feature vector for each user input. These feature vectors will be further examined by the Guardian network to identify potential attackers. Since we do not know whether the new user is under attack and which one of the feature vectors may be from an attacker, we randomly pair the $m$ 512-dimensional feature vectors to create $m$ 1024-dimensional interleaved embeddings as the input for the Guardian network. This will yield $m$ prediction results. The interleaved embedding is obtained from  one attacker's feature vector and one victim's feature vector, or from both normal user's feature vectors. In order to minimize the prediction uncertainty, we employ the KNN (K nearest neighbor) classifier to aggregate the prediction results to produce the binary decision: normal or attacked.  Specifically, during the training of the Guardian network, $m$  1024-dimensional interleaved feature vectors  are used for each user account to produce $m$ probability values. The $m$ probabilities are treated as a $m$-dimensional point for each user. Similarly, the $m$ probability values obtained from the new user registration are also represented as a $m$-dimensional point (denoted as $Pt_{new}$). Then, we search the current dataset to find the top K nearest points to $Pt_{new}$.  If more than $\frac{K}{2}$ of the nearest neighbors  are from the normal accounts, the new user will be labeled as normal. Otherwise, the new user will be considered under attack.

Figure \ref{cnn_guardian} summarizes the complete process of the defense mechanism which consists of four main phases:
\begin{enumerate}
\item {\bf Generating Unbiased Training Data}: We train multiple Deep Speaker models to obtain a balanced set of normal accounts and attacked accounts without affecting the realistic settings in the real world applications where attacked accounts typically exist in a small percentage.

\item {\bf Input Augmentation}: We propose a new way of input augmentation by interleaving feature vector pairs from the same accounts. The augmented input data provides more knowledge to better train  the  deep neural network.

\item {\bf Deep Learning}: We design a convolutional network  to uncover the hidden differences in the feature vectors from attacked and normal accounts.

\item {\bf Prediction Aggregation}: We leverage the power of ensemble learning and calculate multiple prediction results for each given user. We  aggregate the prediction result using KNN to reduce the uncertainty in the prediction and enhance the overall prediction result.

\end{enumerate}

\section{Experimental Studies}

In the experiments, we use two datasets: LibriSpeech \cite{LibriSpeech}, VoxCeleb \cite{VoxCeleb1, VoxCeleb2, VoxCeleb3}. The LibriSpeech and VoxCeleb datasets contain audio files from 2338 and 7245 people, respectively. More statistics can be found in Table \ref{libri_dataset_summary} and \ref{vox_dataset_summary}. For LibriSpeech, we combined the "train-clean-100" and "train-clean-360" partitions for training, and used the "train-other-500" partition for evaluation. All LibriSpeech partitions were trimmed down to contain 10 audio files per user; Both systems were also trained with the VoxCeleb2 dataset, and evaluated with VoxCeleb1. For Deep Speaker, we again trimmed the two VoxCeleb datasets to contain 10 files per user; however, for the VGG Model, we kept the original VoxCeleb2 dataset for training, and the trimmed VoxCeleb1 dataset for evaluation. 10 Deep Speaker models are trained by varying the weight initialization and the set of users being attacked. We evaluate the scenarios when the percentages of attacked users are 5\% or 10\%. For the Guardian network, 10 interleaved feature vectors are generated for each account and the value $K$ is set to 11 for the KNN classifier. To choose the optimal $K$ value, we first plot the curve of the error rate and $K$ with $K$ varying from 1 to 30. Then, according to the graph, we select the $K$ value with the projected minimum error rate.

We compared our Guardian network with SVM, KNN, and a 14-layer fully connected (FC) network.  The SVM and KNN represent the existing outlier-detection based defense mechanisms. The fully connected network resembles the latest defense mechanism \cite{Facial_Authentication} proposed for data poisoning attacks against facial authentication systems. The models used for comparison do not adopt any of  our proposed techniques including  bias reduction,  input augmentation and result aggregation. These models are directly connected to the Deep Speaker and take 512-dimensional feature vectors as input for training and testing.

All the experiments were conducted on a computer with Intel i9-10900X CPU@3.7GHz, NVIDIA GeForce RTX 3090 GPU, and 64GBs of memory. Our Guardian model takes about 25 minutes for training. During evaluation mode, the Deep Speaker model takes 0.954 seconds to register a user, and the Guardian model takes 0.243 seconds to validate a user. In what follows, we focus on evaluating its effectiveness in terms of prediction accuracy, recall, and false positive rate (FPR) as defined in the following equations.

\begin{equation}\nonumber
Accuracy =  \frac{Correctly~~Predicted~~ User~~ Types}{Total~~ Number~~ of~~ New ~~ Users}
\end{equation}

\begin{equation} \nonumber
Recall = \frac{Correctly~~Predicted~~ Attacked~~~ Accounts}{Total~~ Number~~ of~~ Attacked~~ Accounts}
\end{equation}

\begin{equation} \nonumber
FPR = \frac{Incorrectly~~Predicted~~ Normal~~~ Accounts}{Total~~ Number~~ of~~ Normal~~ Accounts}
\end{equation}

\subsection{Comparison with Existing Defense Mechanisms}\label{Varying_the_Percentage_of_Attackers}

At the beginning, we aim to compare the performance of our Guardian network with defense mechanisms that employ conventional machine learning algorithms and the latest data poisoning defense mechanism that used fully connected layers (denoted as "FC" in the figures) \cite{Facial_Authentication}. Table \ref{exp:5-10} shows the attack detection accuracy, recall and false positive rate when there are 5\%  and 10\% of poisoned user accounts.  In both cases, we observe that our proposed Guardian network has achieved around 95\% detection accuracy and recall, whereas other approaches have less than 60\% accuracy and lower recall. The recall of the Guardian is also much higher than the other three approaches. These can be attributed to the series of techniques adopted by the Guardian network.  The results clearly demonstrate the significant benefits of bias reduction, input augmentation, convolutional layers and ensemble learning.



\begin{table}[!ht]
  \renewcommand{\arraystretch}{1.3}
  \caption{Datasets with 5\% and 10\% Attacker Accounts}
  \label{exp:5-10}
  \centering
  \begin{tabular}{c||c|c|c|l}
    \hline
                   Poisoning Ratio & Method & Accuracy & Recall & FPR\\
    \hline\hline
    5\% Attacker Accounts   & SVM       & 0.543  & 0.754  & 0.468\\
                            & KNN       & 0.520  & 0.583  & 0.483\\
                            & FC        & 0.639  & 0.804  & 0.370\\
                            & \textbf{Guardian}  & \textbf{0.944}  & \textbf{0.952}  & \textbf{0.057}\\
    \hline
    10\% Attacker Accounts  & SVM       & 0.488  & 0.452  & 0.508\\
                            & KNN       & 0.520  & 0.528  & 0.481\\
                            & FC        & 0.529  & 0.588  & 0.478\\
                            & \textbf{Guardian}  & \textbf{0.939}  & \textbf{0.910}  & \textbf{0.058} \\
    \hline
  \end{tabular}
  \end{table}

In addition, we observe that the recall of the Guardian network drops slightly but is still around 90\% when the percentage of attacked accounts is doubled. The SVM, KNN and fully connected network have been affected much more than the Guardian network. Specifically, the recall of SVM has been cut by almost half, and the recall of KNN and fully connected network drop below 60\%. This is likely because the increased number of attacked accounts makes the Deep Speaker model fit the attacked accounts better; as a result, the resulting feature vectors from the attacked accounts become even harder to distinguish from normal accounts. However, the experimental results show that our proposed Guardian network is still quite robust with the increase of the amount of attacks. We also argue that it is unlikely that  attackers would have  compromised a large percentage of user accounts (e.g., $>50\%$), which not only requires the attackers to devote significant resources but also increases their chance of being captured during the network interception attack.




\subsection{Effect of Bias Reduction}


In this round of experiments, we are interested in examining the effect of our bias reduction technique alone. Table \ref{exp:bias} shows the detection accuracy and recall with and without the bias reduction technique. The percentage of poisoned accounts is set to 5\%. The version without bias collects training data from only one Deep Speaker model and has the ratio of the attacked accounts to the normal accounts as 1:20. The version with bias reduction  used 4 Deep Speaker models and increased the ratio of the attacked accounts to the normal accounts to 1: 5.

\begin{table}[ht]
  \renewcommand{\arraystretch}{1.3}
  \caption{Effect of Bias Reduction}
  \label{exp:bias}
  \centering
  \begin{tabular}{c||c|c|l}
    \hline
    Bias Reduction  & Accuracy  & Recall  & FPR\\
    \hline\hline
    No              & 0.960     & 0.266   & 0.006\\
    Yes             & 0.944     & 0.952   & 0.057\\
    \hline
\end{tabular}
\end{table}

Observe that the versions with and without the bias reduction technique have similarly high overall accuracy. However, the recall of the version without the bias reduction is extremely low, i.e., less than 30\%. That means without the bias reduction, the model can only detect about 30\% of attacked accounts. The reason the overall accuracy is similar in two versions is because the number of normal accounts is dominant (i.e., 95\%) and the version without bias reduction has no problem identifying normal accounts. After applying the bias reduction technique, the recall has been significantly improved to over 95\%, which indicates the advantages of bias reduction.  In addition, we learn that it is not necessary to train 10 Deep Speaker models to reach 50-50 ratio of attacked and normal accounts in the training samples. The Guardian network already yields satisfactory performance without a fully balanced training dataset, which reduces the training cost.



\subsection{Effect of Input Augmentation}\label{Effect_of_Input_Augmentation}



We also evaluate the effectiveness of our proposed input augmentation,  which interleaves a pair of feature vectors from the same account. Specifically, we modify the Guardian network to take the  512-dimensional feature vectors generated by the Deep Speaker model as input. We still  keep the bias reduction and ensemble learning techniques for the 512-dimensional Guardian network so as to single out the effect of the input augmentation. We compare this modified Guardian network with the Guardian network that has the input augmentation and takes 1024-dimensional feature vectors.



Table \ref{exp:input} reports the performance of these two networks. We can observe that a significant performance improvement has been achieved by the input augmentation technique for both the accuracy and recall. Specifically, the accuracy has been increased from 66\% to 95\%, and the recall has been increased from 35\% to 95\%. This is because when 512-dimensional feature vectors are used, the convolutional layers only study the spatial relationship in a single vector. The interleaved 1024-dimensional feature vectors give the  convolutional layers an opportunity to compare the features from the attacker and victim dimension by dimension, and hence lead to better classification capabilities.

\begin{table}[ht]
  \renewcommand{\arraystretch}{1.3}
  \caption{Effect of Input Augmentation}
  \label{exp:input}
  \centering
  \begin{tabular}{c||c|c|l}
    \hline
    Input Type  & Accuracy  & Recall    & FPR\\
    \hline\hline
    512-D (Original)        & 0.663     & 0.350     & 0.321\\
    1024-D (Augmented)      & 0.944     & 0.952     & 0.057\\
    \hline
\end{tabular}
\end{table}

\subsection{Effect of Ensemble Learning}
In Table \ref{exp:ens}, we show the performance of  two versions of the Guardian network with and without ensemble learning. The one without ensemble learning conducts only one prediction by taking a single 1024-dimensional feature vector from a randomly selected pair of 512-dimensional feature vectors from a new user account. The one with the ensemble learning conducts 10 predictions from  10 randomly selected pairs of the 512-dimensional feature vectors from the same user account.

From the figure, we can see the improvements on both accuracy and recall after adopting the ensemble learning technique. The  increase in recall is much more significant than the accuracy. The reason is the following. As aforementioned, we do not know which input from the new user has been poisoned. The interleaved feature vectors may not include exactly one vector from the attacker and one from the victim. The use of ensemble learning helps select the most confident predictions and hence can identify most of the attacked accounts. On the other hand, the interleaved feature vectors from the normal accounts are always from the original users. Hence, the ensemble learning does not have much impact on the normal accounts, and the high detection accuracy of the normal accounts contribute to the overall accuracy of the version without the ensemble learning.


\begin{table}[!t]
  \renewcommand{\arraystretch}{1.3}
  \caption{Effect of Ensemble Learning}
  \label{exp:ens}
  \centering
  \begin{tabular}{c||c|c|l}
    \hline
    Ensemble Learning & Accuracy  & Recall      & FPR\\
    \hline\hline
    No                & 0.866     & 0.458   & 0.113\\
    Yes               & 0.944     & 0.952   & 0.057\\
    \hline
\end{tabular}
\end{table}

\subsection{Effect of Attackers' Voices}\label{Effect_of_Attackers_Voices}

In the previous experiments, both the victims and attackers are chosen randomly, which means the voices of the victim and the attacker could be relatively similar (e.g., of the same gender) or very different. The attack has been successful in both cases. We are interested in finding out if an attacker who has a similar voice as the victim would impose more challenges on Guardian. In this experiment, we use the Deep Speaker and Guardian models trained for Section \ref{Effect_of_Input_Augmentation}. Then, we test Guardian against two kinds of attackers. Specifically, we select attackers of the same gender as the victims to simulate the similar voice scenario. Table \ref{exp:gender} compares the detection accuracy, recall and false positive rate (FPR) of Guardian under different attack scenarios.  
We can observe that the Guardian’s performance is not affected by the gender between attacker and victim’s voices.

\begin{table}[ht]
  \renewcommand{\arraystretch}{1.3}
  \caption{Effect of Attacker's Gender}
  \label{exp:gender}
  \centering
  \begin{tabular}{c||c|c|l}
    \hline
    Attacker's Voices & Accuracy  & Recall  & FPR\\
    \hline\hline
    Mixed             & 0.944     & 0.952   & 0.057\\
    Same Gender       & 0.942     & 0.967   & 0.059\\
    Different Gender  & 0.942     & 0.976   & 0.060\\
    \hline
\end{tabular}
\end{table}

Next,  we examine more challenging scenarios when the attackers' voices are extremely similar to the victims based on two commonly used similarity metrics: Consine similarity and  clustering metrics. Such attackers are chosen as follows:
\begin{enumerate}
	\item {\bf Cosine Similarity}: Each user account has 10 unique voice files; through the Deep Speaker model, we can obtain 10 corresponding feature vectors. We then calculate the cosine similarities between each pair of feature vectors, resulting in a total of 45 unique comparisons. Finally, by calculating the mean value of these 45 pairs, we can obtain a single cosine similarity score that represents the overall similarity of the 10 voice files. Based on the results,  a threshold for cosine similarity scores is established to allow  majority (95\%) of normal accounts to be classified as acceptable. If the cosine similarity score between an attacker and the victim is above this threshold, such attacker is considered to have extremely similar voices to the victim and added to the test dataset for our Guardian model to check. 
	
	\item {\bf Clustering Metrics (LDA)}: We employ the clustering metrics and within-class scatter matrices concept to define ``similar voices”. More specifically, we use Linear Discriminant Analysis (LDA) to compress the original 512-dimensional feature vector into a smaller 150-dimensional feature vector, the same dimension as the x-vector model \cite{x-vector}. We then calculate the concentration of within-class feature vectors using the following formula:
	
	\begin{equation}\label{eq:lda_mean}
		m_{i} = \frac{1}{N}\sum_{x \subset User_{i}}^{N} x_{k}
	\end{equation}
	
	\begin{equation}\label{eq:lda_sw}
		S_{i} = \sum_{x\subset User_{i}}^{N} (x-m_{i})(x-m_{i})^{T}
	\end{equation}
	
	\begin{equation}\label{eq:concentration_trace}
		ConcentrationScore_{User_{i}} = Trace(S_{i})
	\end{equation}
	
	A threshold for concentration scores is established to allow the majority (95\%) of normal accounts to be classified as acceptable. If an attacker 's concentration score is below this threshold, the attacker is added to the test dataset for the Guardian model to check. 
	
\end{enumerate} 

Table \ref{exp:similar_attacked_account} shows the percentage of such extremely similar attackers that can be detected by our Guardian model.  Observe that among all the attackers which can evade the detection using the cosine similarity threshold, 27\% can still be captured by Guardian; among all the attackers whose concentration scores to the victims are below the clustering metric threshold, around 60\% can still be captured by Guardian. This results indicate the superiority of our Guardian model to approaches that applies simple similarity metrics. It also points out  future research directions on further enhancing Guardian model on detecting such extremely similar voices.

\begin{table}[ht]
\renewcommand{\arraystretch}{1.3}
\caption{Detecting Attackers with Extremely Similar Voices to the Victims}
\label{exp:similar_attacked_account}
\centering
\begin{tabular}{c||c}
	\hline
	Similarity Metrics &  Percentage of Attackers Detected\\
	\hline\hline
	Cosine Similarity     &  27\% \\
	Clustering Metrics    &  60\% \\
	\hline
\end{tabular}
\end{table}

\subsection{Effect of Datasets}

In this round of experiments, we are interested in learning how audio files with noises would affect the performance of Guardian. It is worth noting that for most of voice authentication, audio files are expected to contain less noises. This experiment just aims to test the limit of our approach.

Table \ref{exp:transfer} shows the performance Guardian when detecting poisoned Deep Speaker model  that is trained on LibriSpeech and VoxCeleb, respectively.   The VoxCeleb dataset contains lots of background noises compared to LibriSpeech, and the Guardian model's detection accuracy in VoxCeleb dataset is lower than that in the LibriSpeech dataset. This is because the voice recognition accuracy of the Deep Speaker with respect to VoxCeleb dataset is already lower than that with respect to LibriSpeech as shown earlier in Section \ref{sec:voice_recognition_systems}.  As a result, this low voice recognition accuracy (88\%) of the Deep Speaker affects the detection accuracy of the Guardian model. To sum up,  the Guardian model will perform well when the corresponding voice recognition model performs well.

\begin{table}[!t]
  \renewcommand{\arraystretch}{1.3}
  \caption{Model Transferability}
  \label{exp:transfer}
  \centering
  \begin{tabular}{c||c|c|l}
    \hline
    Dataset                                 & Accuracy  & Recall    & FPR\\
    \hline\hline
    Deep Speaker with LibriSpeech                  & 0.944     & 0.952     & 0.057\\
    Deep Speaker with VoxCeleb              & 0.863     & 0.831     & 0.135\\
    \hline
\end{tabular}
\end{table}


\subsection{Detecting Pure Man-in-the-Middle Attack}

Finally, we examine the capability of our Guardian model  in terms of detecting pure man-in-the-middle (MiTM) attacks. As shown in Table 4 in Section 3.3, pure MiTM attacks have success attack rate around 88\%.   Table \ref{exp:without_data_poisoning} shows the performance of our Guardian model  for the poisoned and clean authentication models. We observed that the Guardian model could achieve even better performance when the voice authentication model has not been poisoned. This also indicates that it would be easier for attackers to impersonate the victims if the attackers can poison the model during the training phase. On the other hand, poisoned data leads to bad feature vectors which imposes greater challenges to the Guardian model due to the less clear differences between the two feature vectors as compared to those generated by a clean model. Applying the Guardian to the clean model eliminates the adverse effects of data poisoning attacks, which explains the better detection accuracy in this experiment.

\begin{table}[ht]
  \renewcommand{\arraystretch}{1.3}
  \caption{Success Rate of Detecting Pure Man-in-the-Middle Attacks}
  \label{exp:without_data_poisoning}
  \centering
  \begin{tabular}{c||c|c|l}
    \hline
    Dataset                                 & Accuracy  & Recall    & FPR\\
    \hline\hline
    Poisoned Deep Speaker (LibriSpeech)     & 0.944     & 0.952     & 0.057\\
    Clean Deep Speaker (LibriSpeech)        & 0.970     & 0.981     & 0.031\\
    Poisoned VGG (LibriSpeech)              & 0.953     & 0.939     & 0.046\\
    Clean VGG (LibriSpeech)                 & 0.963     & 0.939     & 0.036\\
    \hline
\end{tabular}
\end{table}

\section{Security Analysis}

We now  analyze potential attacks to our Guardian model. The first scenario that may seem to  challenge the effectiveness of the Guardian model is if the attacker has a similar voice to the victim. The similarity here simply refers to how the voice sounds to the human ears. Fortunately, the voice recognition models have much better voice identification capabilities than humans. They already have very high recognition accuracy among a large number of users whereby users of similar voices inevitably exist. In our experiments (Section \ref{Effect_of_Attackers_Voices}), we also simulated such a scenario by selecting same-gender attacker and victims since their voices would be more similar than those from different genders. The experimental results show that our Guardian model performs similarly well in both cases. In the future, to further investigate the performance of the Guardian, we plan to train our Guardian model with ``similar voices" instead of randomly selected attacked users.

\begin{table}[ht]
  \renewcommand{\arraystretch}{1.3}
  \caption{Voice Recognition Accuracy Under Fake Voice Attacks}
  \label{exp:deepspeaker_fake}
  \centering
  \begin{tabular}{c||c|l}
    \hline
    Attack Name     & Overall Accuracy & Recognition Accuracy\\
    \hline\hline
    No Fake Voice   & 0.961           & NA\\
    5\% Fake Voice  & 0.956          & 0.956(Victim)/0.953(Attacker)\\
  \hline
\end{tabular}
\end{table}

Another scenario is when the attacker knows the existence and architecture of our Guardian model and attempts to take advantage of that. The attackers may try to use  some existing Text-to-Speech (TTS) techniques to generate voice files that mimic the victim's voices so as to fool the voice authentication model and the Guardian model. However, even if the fake voices are successfully generated to fool the voice authentication model and the Guardian model, they will not be able to escape from the fake voice detector which is good at distinguishing fake voices from authentic human voices. We have conducted the following experiments to validate our conjecture. First, we used a well-known repository called Real-Time Voice Cloning \cite{RealTimeVoiceCloning} to generate  fake voices. This repository is an implementation of SV2TTS \cite{sv2tts}. There are three different neural network structures in this system, and we use the pre-trained models they provided during our experiments. We chose 1166 users in LibriSpeech \cite{LibriSpeech} dataset. Keeping the same settings as that in the previous experiments, each user has 10 original voice files, and 5\% of user accounts (i.e., victims) contain fake voice files. Each victim account has 5 original voice files and 5 fake voice files generated by the Real-Time Voice Cloning system. Table \ref{exp:deepspeaker_fake} shows the accuracy of voice recognition before and after the data poisoning attack. Before the attack, the overall voice recognition accuracy is around 0.991. After some user accounts being injected with fake voices, the recognition accuracy of both normal accounts and victim accounts still stay at a very high level as shown in the table. That means the fake voices have successfully fooled the voice recognition model. Next, we check how the Guardian model reacts to the fake voices. Table \ref{exp:guardian_fake} shows the detection accuracy and recall. From the low recall rate, we can see that the Guardian model misses the majority of fake voices.


\begin{table}[ht]
  \renewcommand{\arraystretch}{1.3}
  \caption{Guardian with 5\% Fake Voice Users}
  \label{exp:guardian_fake}
  \centering
  \begin{tabular}{c||c|c|l}
    \hline
    Poisoning Ratio &Accuracy&Recall&FPR\\
    \hline\hline
    5\% Fake Voice & 0.953 & 0.136 & 0.005\\
  \hline
\end{tabular}
\end{table}

The results from Table \ref{exp:deepspeaker_fake} and \ref{exp:guardian_fake} indicate that both the voice recognition model and the Guardian model are unaware of the fake voice attack. However, there is an easy way for the service providers to filter out such fake voices by employing the existing fake voice detectors such as Deep Sonar \cite{DeepSonar}. This detector is based on monitoring neuron behaviors of the voice recognition system  to discern synthesized fake voices.
In order to observe the performance of Deep Sonar in our system, we trained a brand new Deep Sonar network for the Deep Speaker using the LibriSpeech dataset. Table \ref{exp:deepsonar} shows  the high detection accuracy and recall achieved by Deep Sonar with respect to the fake voice attack. This means Deep Sonar is a very effective tool to detect fake voices and is powerful enough to defend against the fake voice attack. 

Note that fake voice detectors are a good complement to the overall voice authentication system, but cannot replace the function of the Guardian model  since fake voice detectors are only versed at identifying  manipulated voices. The second row in Table \ref{exp:deepsonar}  shows nearly zero recall rate with respect to poisoning attacks using human voices. In other words,  Deep Sonar is not able to filter out  authentic human voices when the attacker directly injected their real voices in the victim's account like the targeted data poisoning attack in our case.

\begin{table}
  \renewcommand{\arraystretch}{1.3}
  \caption{DeepSonar under Different Types of Attacks}
  \label{exp:deepsonar}
  \centering
  \begin{tabular}{c||c|c|l}
    \hline
    Attack Name&Accuracy&Recall&FPR\\
    \hline\hline
    5\% Fake Voice Attack & 0.941 & 0.939 & 0.059\\
    5\% Poisoning Attack & 0.891 & 0.05 & 0.065\\
  \hline
\end{tabular}
\end{table}

We also consider the scenario when attackers attempt to apply  the idea of Generative Adversarial Networks (GAN) \cite{GAN} to produce  voice files that can fool the voice authentication model, the fake voice detector, and the Guardian model. However,  it would be extremely challenging to implement such an attack. Recall that both the fake voice detector and the Guardian model hide behind the voice recognition model. As the attacker does not have the specific parameters of any of these three models, they may train their own system  with the same structures. Since all the three models are deep neural networks, the back propagation process is very complicated. In our trial with known system parameters, we are still not able to make such training converge. Even if the attackers managed to complete the local training, it is unclear if the locally generated fake voices will be sufficiently effective to fool the real models used by their targeted service providers.

Finally, we consider limitations around the specific attack and threat model. While our threat model includes a data poisoning attack and a MiTM attack, since this area of research is still relatively new, other sophisticated attacks  may emerge as well.  For example, \cite{curie} analyzes data poisoning attacks in situations where the model is periodically re-trained whereas in our work, the voice authentication system is assumed not to be retrained any more.  Additionally, our Guardian model is designed to protect CNN-based voice recognition models. It is worth studying whether Guardian may also be effective for other architectures like transformers and wav2vec \cite{schneider2019wav2vec}.


\section{Conclusion}
In this paper, we investigate a targeted data poisoning attack that allows the attacker to impersonate a legitimate user via voice authentication.  We propose a novel CNN-based discriminator called Guardian to help  distinguish the attacked accounts from normal accounts. We design a series of advanced techniques for the Guardian network to obtain balanced training samples and augmented input feature vectors, which significantly improves the Guardian network's effectiveness. Our experimental results demonstrate that the Guardian network achieves around 95\% detection accuracy while existing defense mechanisms only yield  60\% accuracy. In the future, we are interested in exploring new attack scenarios and the effectiveness of our Guardian model against non CNN-based voice authentication models.


%
\bibliographystyle{IEEEtran}
\bibliography{IEEE}

%

\begin{IEEEbiography}[{\includegraphics[width=1in,height=1.25in,clip,keepaspectratio]{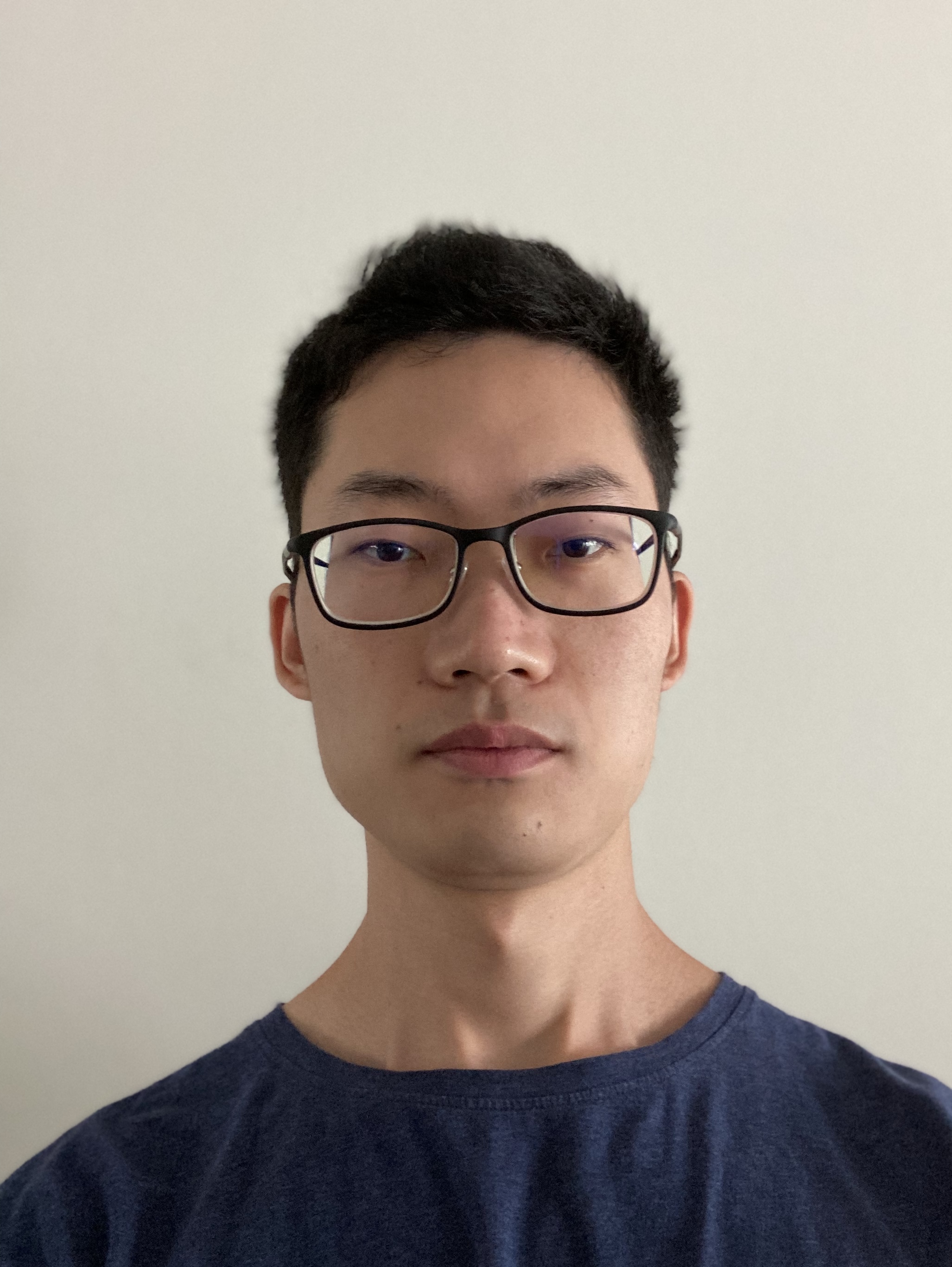}}]{Ke Li} received his B.S. degree in 2019 from the East China University of Science and Technology, Shanghai, China. He is currently working toward a PhD degree in Computer Science at Vanderbilt University. His research interests include machine learning, voice authentication, and adversarial attacks.
\end{IEEEbiography}

\begin{IEEEbiography}[{\includegraphics[width=1.15in,height=1.5in,angle=-90,clip,keepaspectratio]{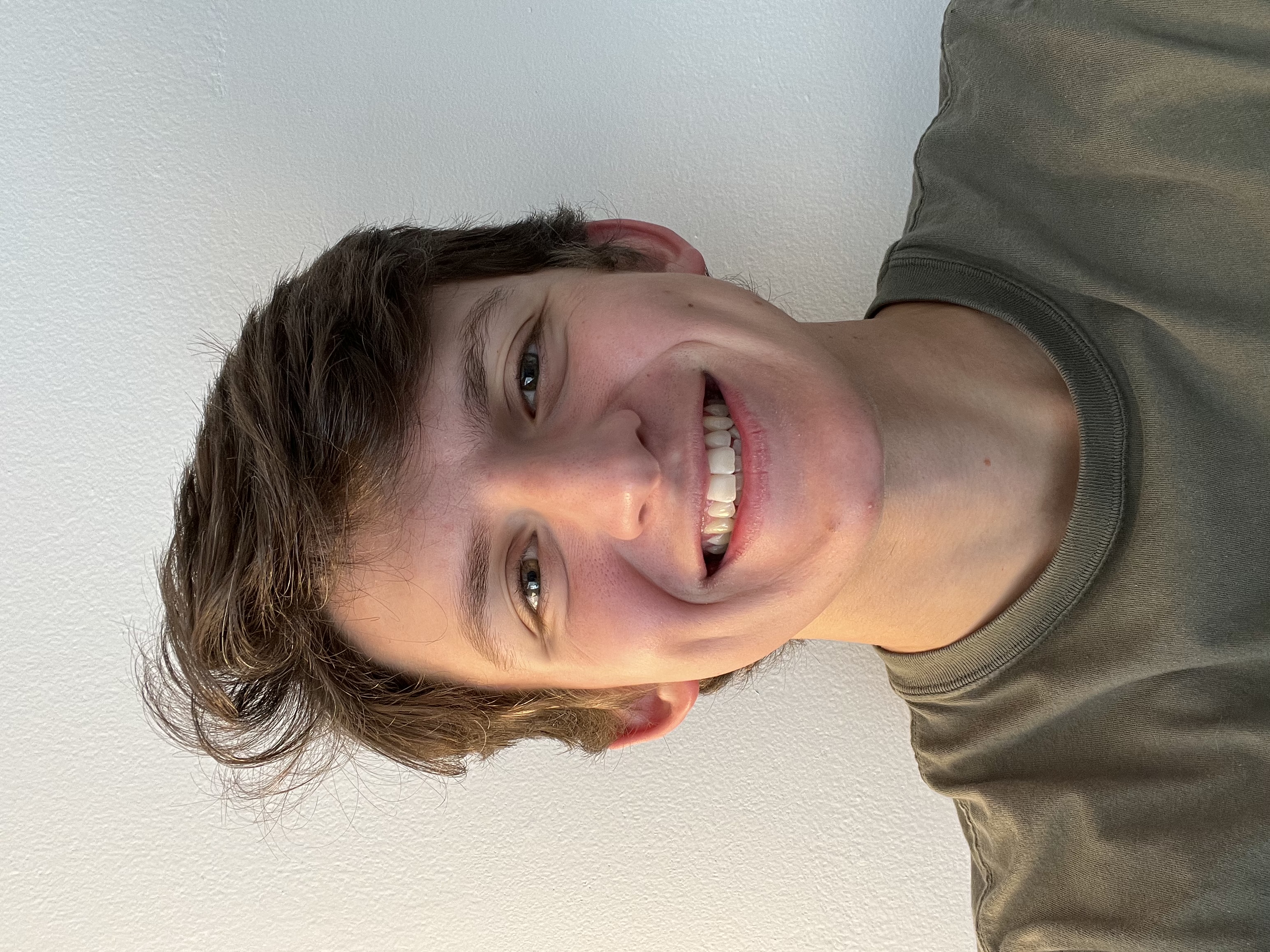}}]{Cameron Baird} is currently a PhD student in the Department of Computer Science at Vanderbilt University. He received the bachelor's degree in computer science from the University of Missouri-Columbia in 2022. His research interests include adversarial attacks against machine learning based systems.
\end{IEEEbiography}

\begin{IEEEbiography}[{\includegraphics[width=1in,height=1.25in,clip,keepaspectratio]{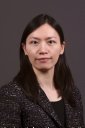}}]{Dan Lin}  is currently a professor in Department of Computer Science and Director of I-Privacy Lab at Vanderbilt University. She received her PhD degree in Computer Science from  National University of Singapore in 2007, and was a postdoctoral research associate at Purdue University for two years.  Her research interests cover many areas in the fields of information security and database systems.
\end{IEEEbiography}




\end{document}